\documentclass{aa}
\usepackage{graphicx}
\usepackage{txfonts}
\newcommand{\prodimo}{P{\tiny RO}D{\tiny I}M{\tiny O}\;}

\usepackage{natbib}
\bibpunct{(}{)}{;}{a}{}{,} % to follow the A&A style
\usepackage{color}
\usepackage[colorlinks=true,linkcolor=black,citecolor=blue,urlcolor=blue,draft=false]{hyperref}% To add links in your PDF file, use the package "hyperref"
\makeatletter
\renewcommand*\aa@pageof{, page \thepage{} of \pageref*{LastPage}}
\makeatother

\begin{document} 

   \title{Observing the gas component of circumplanetary disks around wide-orbit planet-mass companions in the (sub)mm regime}

  \author{Ch. Rab\inst{1}  
  \and I. Kamp\inst{1}
  \and C. Ginski\inst{2,3}
  \and N. Oberg\inst{1}
  \and G. A. Muro-Arena\inst{2}
  \and C. Dominik\inst{2}
  \and L. B. F. M. Waters\inst{4,2}
  \and W.-F. Thi\inst{5}
  \and P. Woitke\inst{6}}

  \institute{Kapteyn Astronomical Institute, University of Groningen, P.O. Box 800, 9700 AV Groningen, The Netherlands \email{rab@astro.rug.nl}
  \and Anton Pannekoek Institute for Astronomy, University of Amsterdam, Science Park 904, 1098 XH Amsterdam, The Netherlands
  \and Leiden Observatory, Leiden University, PO Box 9513, 2300 RA Leiden, The Netherlands
  \and  SRON Netherlands Institute for Space Research, Sorbonnelaan 2, 3584 CA Utrecht, The Netherlands
  \and Max-Planck-Institut f\"ur extraterrestrische Physik, Giessenbachstrasse 1, 85748 Garching, Germany
  \and SUPA, School of Physics \& Astronomy, University of St. Andrews, North Haugh, St. Andrews KY16 9SS, UK
  } 

   \date{Received December 17 2018 / Accepted February 6 2019}

% \abstract{}{}{}{}{} 
% 5 {} token are mandatory
 
  \abstract
  % context heading (optional)
  % {} leave it empty if necessary  
   {Several detections of wide-orbit planet-mass/substellar companions around young solar-like stars  were reported in the last decade. The origin of those possible planets is still unclear, but accretion tracers and VLT/SPHERE observations indicate that they are surrounded by circumplanetary material or even a circumplanetary disk (CPD).}
  % aims heading (mandatory)
   {We want to investigate if the gas component of disks around wide-orbit companions is detectable with current (ALMA) and future (ngVLA) (sub)mm telescopes and what constraints such gas observations can provide on the nature of the circumplanetary material and the mass of the companion.}
  % methods heading (mandatory)
   {We applied the radiation thermochemical disk code \prodimo to model the dust and gas component of passive CPDs and produced realistic synthetic observables. We considered different companion properties (mass, luminosity), disk parameters (mass, size, dust properties) and radiative environments (background fields) and compared the resulting synthetic observables to telescope sensitivities and existing dust observations.}
  % results heading (mandatory)
   {The main criterion for a successful detection is the size of the CPD. At a distance of about $150\,\mathrm{pc}$, a CPD with an outer radius of about $10\,\mathrm{au}$ is detectable with ALMA in about six hours in optically thick CO lines. Other aspects, such as the luminosity, disk inclination, and background radiation fields of the companion, are also relevant and should be considered to optimize the observing strategy for detection experiments.}
  % conclusions heading (optional), leave it empty if necessary 
   {For most of the known wide-orbit planet-mass companions, their maximum theoretical disk size of one-third of the Hill radius would be sufficient to allow detection of CO lines. It is therefore feasible to detect their gas disks and constrain the mass of the companion through the kinematic signature. Even in the case of non-detections such observations provide stringent constraints on disk size and gas mass, and this information is crucial for formation theories.}

\titlerunning{Observing the gas component of CPDs in the (sub)mm regime}

   \maketitle
%
%-------------------------------------------------------------------
\section{Introduction}
In the last decade several detections of substellar or planet-mass companions (PMCs; i.e. $M_\mathrm{p}\lesssim 20\,M_\mathrm{J}$) orbiting young \mbox{($\approx 1-10\,\mathrm{Myr}$)} solar-like stars on wide orbits $(a\gtrsim100\,\mathrm{au}$) were reported. Most of those companions show or were even detected via accretion tracers such as $H_\alpha$ or $P_\gamma$ lines (e.g. \citealt{Neuhaeuser2005,Ireland2011,Bowler2011,Zhou2014,Currie2014,Kraus2014,Wu2015,Santamaria-Miranda2018}). This indicates that those objects are likely embedded in circumplanetary material and possibly host a circumplanetary disk (CPD). The origin of those PMCs is still unclear; they might have been formed in protoplanetary disks via core accretion and subsequently scattered towards wider orbits, formed in gravitationally unstable disks, or followed a similar formation pathway as wide binaries (see e.g. \citealt{Boss2006,Vorobyov2013,Stamatellos2015,Rodet2017}). 

Using polarized light observations with SPHERE at the Very Large Telescope (VLT/SPHERE), \mbox{\citet{Ginski2018}} detected a CPD candidate on a wide orbit ($a\approx215\,\mathrm{au})$ around the close-binary system \mbox{CS Cha}. Their measured polarization degree provides strong evidence for the presence of dusty circumplanetary material around the PMC \mbox{CS Cha c} and is also consistent with the presence of a CPD. Their observations indicate that the companion is not embedded in the disk of the primary system, which also seems to be the case for other \mbox{wide-orbit} PMCs \citep{Wu2017}. This makes PMCs ideal targets to hunt for CPDs.

\citet{Wu2017} observed several wide-orbit PMCs with ALMA (Atacama Large Millimeter Array) but did not detect any millimetre continuum emission at the location of the PMCs. These authors concluded that the millimetre emission of the possible CPDs might be optically thick and compact with an outer radius of only \mbox{$r_\mathrm{out}<1000\,R_\mathrm{J}\;(\approx0.5\,\mathrm{au})$} and therefore the CPDs were not detected. Applying similar assumptions, \citet{Wolff2017} derived \mbox{$r_\mathrm{out}<2.9\,\mathrm{au}$} for the potential CPD of the DH Tau b companion from their NOrthern Extended Millimeter Array (NOEMA) continuum non-detections. In contrast to those observations, \citet{Bayo2017} detected a CPD around the free-floating planet-mass object \object{OTS 44} ($M_\mathrm{p}\approx6-17\,M_\mathrm{J}$) with ALMA. Their observed continuum peak flux of $101\,\mathrm{\mu Jy}$ at $233\,\mathrm{GHz}$ is very close to the upper limits of $100-200\,\mathrm{\mu Jy}$ derived by \citet{Wu2017} for their CPD sample. However, the CPD around OTS~44 is unresolved with a beam size of $1{\arcsec}.6 \times 1{\arcsec}.6$ ($r_\mathrm{out}\lesssim 130\,\mathrm{au}$ at a distance of $160\,\mathrm{pc}$).

Although the detection of CPDs in the dust is a very important first step it does not provide constraints on the nature of the companion (i.e. mass) and the properties of the gaseous circumplanetary material. If the gaseous CPD emission can be spectrally and spatially resolved, observed rotation would be a clear sign of the presence of a disk-like structure and would allow us to measure the dynamical mass of the companion.

\citet{MacGregor2017} reported $\mathrm{CO}\,J\!=\!3\!-\!2$ and dust continuum observations of the GQ~Lup system. Their observations indicate that the PMC \mbox{\object{GQ Lup b}} might be still embedded in the disk of the primary, but they did not detect any signature of a CPD nor any disturbance in the Keplerian velocity field of the disk of the primary. These authors concluded that higher spatial resolution and higher sensitivity observations are required to constrain the nature of \mbox{GQ Lup b}. Another interesting example is the companion \mbox{\object{FW Tau c}} \citep{White2001,Kraus2014} in the close-binary system \object{FW~Tau}. \citet{Caceres2015} detected $^{12}\mathrm{CO}\,J\!=\!2\!-\!1$ and continuum emission around \mbox{FW Tau c} with ALMA and derived a companion mass of $M_\mathrm{p}\lesssim35\,M_\mathrm{J}$ but could not exclude higher masses. However, \citet{Wu2017a} used higher spatial and spectral resolution ALMA observations to derive the dynamical mass of the companion and find that \mbox{FW Tau c} is a low-mass star with $M_*\approx0.1\,M_\mathrm{\sun}$ surrounded by a gas disk with $r\approx140\,\mathrm{au}$ (but the dust disk remains unresolved). Interestingly no disk around the central binary system was detected. The example of FW~Tau shows the power of (sub)millimetre (hereafter (sub)mm) gas observations to unambiguously confirm the nature of wide-orbit companions.

Detecting CPDs in the (sub)mm regime is certainly challenging as the already performed dust observations have shown. For the gas component it is even more challenging because of the narrow bandwidths required to detect spectral lines. However, similar to disks around young solar-like stars the apparent gas disk size might be significantly larger than for the dust \citep[e.g.][]{Ansdell2018,Facchini2017}. Furthermore \citet{Zhu2018} have argued that CPDs could be strongly dust depleted owing to efficient radial migration of millimetre-sized dust grains, if there is no mechanism to stop the migration \citep{Pinilla2013}.

So far only a few theoretical studies investigated the detectability of CPDs in the gas. \citet{Shabram2013} analytically estimated CO line fluxes for their radiation-hydrodynamics model of wide-orbit CPDs and found that those still embedded CPDs should be easily detectable with ALMA.  \citet{Perez2015a,Perez2018} studied the detectability of planets and their CPDs via their imprint in the gas dynamics of the circumstellar disk. \citet{Pinte2018} indeed observed signatures of an embedded planet in the \mbox{HD 163296} at an orbit of $a\approx260\,\mathrm{au}$, but could not detect a CPD due to limited spatial resolution. However, for PMCs still embedded in their parent protoplanetary disk, inferring properties of a possible CPD is extremely challenging.

In this work we investigate the possibility to detect CPDs around PMCs on wide orbits with (sub)mm telescopes such as ALMA and the future next generation Very Large Array (ngVLA) and aim to answer the question if CPDs are actually easier to detect in the gas than in the dust at long wavelengths. For this, we use the radiation thermochemical disk model \prodimo (PROtoplanetary DIsk MOdel) to self-consistently model the gas and dust component of CPDs assuming that they are already separated from the disk of their host star. 

In Sect.~\ref{sec:methods}, we describe our model for the CPD and our procedure to produce synthetic observables. Our results are presented in Sect.~\ref{sec:results}, in which we discuss the impact of disk structure, dust properties, and the radiative environment on the resulting line and dust emission. In Sect.~\ref{sec:discussion}, we discuss the challenges to detect those CPDs and what we can learn from deep observations in the (sub)mm regime. We conclude with a summary of our main findings in Sect.~\ref{sec:conclusions}.
%-------------------------------------------------------------------
\section{Methods}
\label{sec:methods}
The knowledge about CPDs around wide-orbit PMCs is still limited. Most theoretical studies focussed on early evolutionary stages during which the CPD is still embedded in their parent protoplanetary disks \citep[e.g.][]{Ayliffe2009,Shabram2013,Perez2015a,Szulagyi2018} and most of these  works only considered orbits of tens of au. In this stage the accretion of material from the surrounding protoplanetary disk likely has a significant impact on the physical structure of the CPD. \citet{Zhu2018} have considered, in their analytical models, viscous heating and irradiation dominated CPDs for planets on close orbits. These authors found that for the detectability of the dust continuum it is not relevant what heating process is the dominant one.

For our modelling we neglect any possible impact of the primary's disk on the CPD. We implicitly assume that the companion and its CPD are either already completely separated from the disk of the host star or were not formed in the protoplanetary disk at all (see Sections \ref{sec:formationscenario} and \ref{sec:embeddedpmc}). Furthermore, we assume that the main heating source of the disk is the radiation of the companion and we neglect viscous heating (see \citealt{Zhu2018}) and the radiation of the host star. Neglecting viscous heating is a reasonable simplification considering the low measured accretion rates of the known wide-orbit PMCs (see Sect~\ref{sec:protoplanet}) and our focus on (sub)mm observations. The wide orbits of known PMCs suggest that the contribution of the stellar radiation to the total irradiation is likely insignificant as for example shown by \citet{Ginski2018} for the \mbox{CS Cha} companion. However, we also present models considering a possible strong stellar or background radiation field. For the gas and dust density structure of the disk, we chose a simple but flexible parametric approach which allows us to explore to some extent the parameter space of the CPD and the PMC properties. 

In Sect.~\ref{sec:methodrefmodel} we describe our reference model for the PMC and its CPD. In Sect.~\ref{sec:rtcmodelling} we briefly discuss the radiation thermochemical disk code \prodimo \citep{Woitke2009a}, which we used to calculate the disk radiation field, gas and dust temperatures, chemical abundances, and synthetic observables (Sect.~\ref{sec:methodsyntheticobs}).
\subsection{Reference model}
\label{sec:methodrefmodel}
In this Section we present our modelling approach for the radiation properties of the PMC and the structure of the disk. We describe in detail our reference model and its parameters. We use this model to discuss several general observational properties of CPDs (e.g. apparent disk radii). However, the reference model should mainly be seen as a starting point as we vary several parameters of the PMC and the CPD model (i.e. masses, luminosities, and dust properties).  
\subsubsection{Planet-mass companion}
\label{sec:protoplanet}
To model the irradiation of the disk by the PMC we need to know the mass $M_\mathrm{p}$, the luminosity $L_\mathrm{p}$,  and the effective temperature $T_\mathrm{p}$. For the known wide-orbit PMCs typical values of about $M_\mathrm{p}\approx10-20\,M_\mathrm{J}$, $L_\mathrm{p}\approx10^{-2} - 10^{-3}\,L_\mathrm{\sun}$, and $T_\mathrm{p}\approx1000-2500\,\mathrm{K}$ are reported \citep[e.g.][]{Wu2017,Ginski2018}. For our reference model we chose $M_\mathrm{p}=20\,M_\mathrm{J}$, $L_\mathrm{p}\approx10^{-2}\,L_\mathrm{\sun}$, and $T_\mathrm{p}=2500\,\mathrm{K}$. As we are mainly interested in the detectability of the CPDs we used values at the upper end of the reported parameter range. For example a higher luminosity makes the disk warmer and therefore easier to detect. We are aware that the properties chosen above do not necessarily describe a planet because, for example, the mass is above the brown-dwarf/deuterium burning limit of $13\,M_\mathrm{\sun}$. However, we also discuss models with lower masses and luminosities that are more appropriate, for example for giant gas planets.

\begin{figure}
\centering
\resizebox{\hsize}{!}{\includegraphics{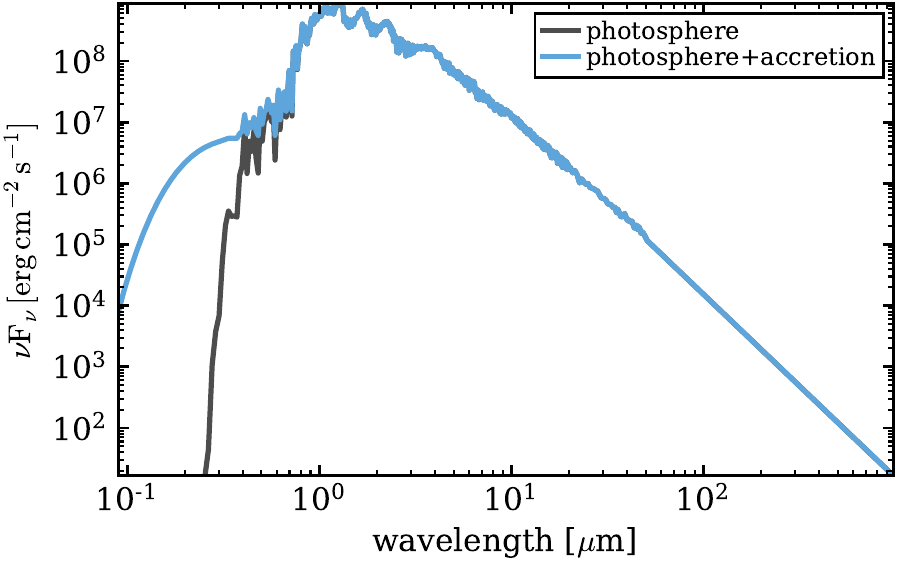}}
\caption{Spectrum for our reference PMC with \mbox{$M_\mathrm{p}=20\,M_\mathrm{J}$}. The solid black line indicates the photospheric (intrinsic) spectrum and the blue line shows the spectrum with the added accretion luminosity.}
\label{fig:starspecJ20}
\end{figure}
Besides the intrinsic luminosity of the planet we also considered accretion luminosity. The reported accretion rates in the literature are in the range $\dot{M}_\mathrm{accr}\approx 10^{-12}-10^{-10}\,M_\mathrm{\sun}\mathrm{yr}^{-1}$ (see \citealt{Wu2017}). We used an accretion luminosity of $L_\mathrm{p}\approx10^{-4}\,L_\mathrm{\sun}$ (i.e. 1\% of the photospheric luminosity); this translates into mass accretion rates of $\dot{M}_\mathrm{accr}\approx 8.9\times10^{-11}\,M_\mathrm{\sun}\mathrm{yr}^{-1}$. The details on how we constructed the planetary input spectrum and calculated the according mass accretion rates are described in Appendix~\ref{sec:planetaryspectra}. In Fig.~\ref{fig:starspecJ20} we show the resulting PMC input spectrum.
\subsubsection{Circumplanetary disk model}
\label{sec:cpdstructure}
As already mentioned, the knowledge of CPD structure of wide-orbit PMCs is limited and their formation mechanism is still unknown. We therefore used as a starting point the reference T~Tauri disk model of \citet{Woitke2016} and scaled the structure properties (i.e. disk mass) according to the masses of PMCs. Motivated by the observations of accretion in the wide-orbit CPDs we assumed that their disk structure is mostly determined by viscous evolution.

\begin{figure}
\centering
\resizebox{\hsize}{!}{\includegraphics{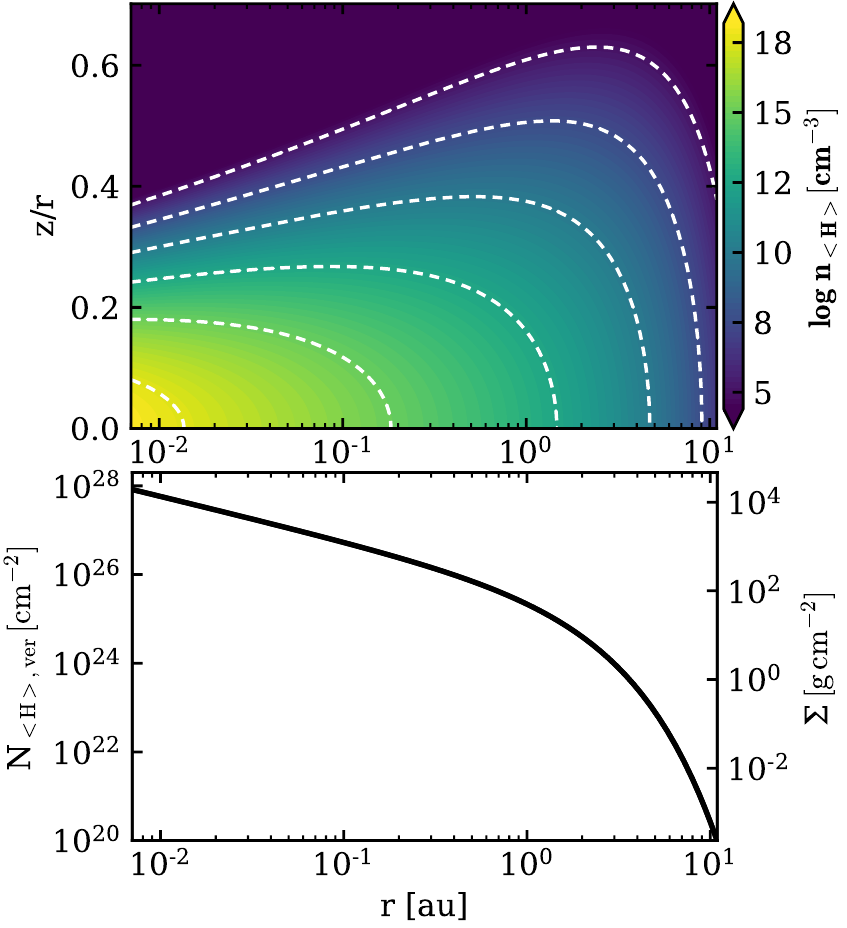}}\\
\caption{Gas disk structure of the reference CPD model. The top panel shows the total hydrogen number density $n_\mathrm{\langle H\rangle}$. The height of the disk $z$ is scaled by the radius $r$. The white dashed contours correspond to the density levels shown in the colour bar. The bottom panel shows the total vertical hydrogen column number density $N_\mathrm{\langle H\rangle,ver}$ as a function of radius; the scale for the surface density $\Sigma$ in $\mathrm{g\,cm^{-2}}$ is given on the right-hand side.}
\label{fig:diskstruc}
\end{figure}
We used a simple parameterized and fixed disk structure \citep[see e.g.][]{Woitke2016}. The density structure of the disk is given by 
\begin{equation}
  \label{eqn:density}
    \rho(r,z)=\frac{\Sigma(r)}{\sqrt{2\pi}\cdot h(r)}\exp\left(-\frac{z^2}{2h(r)^2}\right)\;\;\mathrm{[g\,cm^{-3}]}\;,   
\end{equation}
where the radius $r$ and height $z$ are in cylindrical coordinates.\ The value $\Sigma(r)$ is the disk surface density of the disk and $h(r)$ is the scale height of the disk. For $\Sigma(r)$ we used either a power-law prescription with an exponentially tapered outer edge or a pure power law. The surface density profile is scaled according to the given disk mass $M_\mathrm{d}$. In the tapered outer edge models the outer disk radius $r_\mathrm{out}$ is not a parameter but is defined as the radius where the total vertical hydrogen column density reaches $N_\mathrm{\langle H \rangle,ver}=10^{20}\,\mathrm{cm^{-2}}$ (see Fig.~\ref{fig:diskstruc}). The scale height of the disk is also parameterized via a simple power law.

So far only upper limits for the dust mass of CPDs exist and the gas-to-dust mass ratio and consequently the total disk mass is essentially unknown. We therefore chose for our reference model a disk mass of 1\% of the PMC (similar to T Tauri disks; e.g. \citealt{Andrews2013x}) and assume the canonical dust-to-gas mass ratio of $d/g=0.01$. The resulting dust mass is actually above the so far reported upper limits for CPD dust disk masses (see also Sect.~\ref{sec:results_struc}) derived from ALMA continuum observations (e.g. \citealt{Wu2017,Wolff2017}). However, that does not necessarily mean that our assumed mass for the reference model is in disagreement with the observations (see Sect.~\ref{sec:results_struc}).

An upper limit for the disk size of CPDs comes from the Hill radius of the companion
\begin{equation}
r_\mathrm{Hill}= a\left(\frac{M_\mathrm{p}}{3M_\mathrm{*}}\right)^{1/3}.
\end{equation} 
In this equation $a$ is the semimajor axis of the orbit and $M_\mathrm{p}$ and $M_\mathrm{*}$ are the mass of the companion and the stellar host, respectively. Material outside of the Hill sphere is not orbiting the planet and is therefore not part of the disk. However, analytical and numerical models suggest that CPDs might be truncated at $0.3-0.4\,r_\mathrm{Hill}$ owing to tidal truncation effects and/or the low specific angular momentum of infalling material during their formation \citep[e.g.][]{Quillen1998,Martin2011,Shabram2013}. Although this more strict requirement does not necessarily apply to wide-orbit PMCs, i.e. which likely depends on their formation mechanism, we used $r_\mathrm{Hill}/3$ as a reference quantity for the outer radius of our CPD models.

For the reference model we chose a tapering-off radius of $R_\mathrm{tap}=1 \,\mathrm{au}$, resulting in an outer radius of $r_\mathrm{out}=10.9\,\mathrm{au}$. This value is equal to $r_\mathrm{Hill}/3$ for a PMC with $M_\mathrm{p}=20\,M_\mathrm{J}$ orbiting a solar-mass star at an orbital distance of $a=178\,\mathrm{au}$. We present models with smaller and larger disk sizes and also discuss the different apparent (observed) disk sizes for the dust and gas in detail in Sect.~\ref{sec:res_reference}. We determined the inner radius of our disk models by the dust sublimation radius, where the dust temperature reaches $\approx1500\,\mathrm{K}$.

The density structure and surface density profile for our reference model are shown in Fig.~\ref{fig:diskstruc}; the gas and dust temperature structures are shown in Fig.~\ref{fig:reftemp}. All relevant parameters for our reference model are listed in Table~\ref{table:discmodel}.

\begin{table}
\caption{Main parameters for our reference model.}
\label{table:discmodel}
\centering
\begin{tabular}{l|c|c}
\hline\hline
Quantity & Symbol & Value  \\
\hline
companion mass                          & $M_\mathrm{p}$                    & $20~M_\mathrm{J}$\\
companion effective temp.               & $T_{\mathrm{p}}$                  & 2500~K\\
companion luminosity                    & $L_{\mathrm{p}}$                  & $10^{-2}~L_\sun$\\
\hline
strength of interst. FUV              & $\chi^\mathrm{ISM}$               & 1\tablefootmark{a}\\
\hline
disk gas mass                         & $M_{\mathrm{d}}$                  & $0.2~M_\mathrm{J}$\\
dust/gas mass ratio                   & $d/g$                             & 0.01\\
inner disk radius                     & $R_{\mathrm{in}}$                 & 0.007~au\\
tapering-off radius                   & $R_{\mathrm{tap}}$                & 1~au\\
column density power ind.             & $\epsilon$                        & 1.0\\

reference scale height                & $H(1\;\mathrm{au})$               & 0.1~au\\
flaring power index                   & $\beta$                           & 1.15\\
\hline
min. dust particle radius             & $a_\mathrm{min}$                  & $\mathrm{0.05~\mu m}$\\
max. dust particle radius             & $a_\mathrm{max}$                  & 3 mm\\
dust size dist. power index           & $a_\mathrm{pow}$                  & 3.5\\
max. hollow volume ratio\tablefootmark{b}              & $V_{\mathrm{hollow,max}}$         & 0.8\\
dust composition \tablefootmark{c}                     & {\small Mg$_{0.7}$Fe$_{0.3}$SiO$_3$}  & 60\%\\
(volume fractions)                    & {\small amorph. carbon}                    & 15\%\\
& {\small porosity}                          & 25\%\\
\hline
inclination                           & $i$                    & $45^\circ$    \\
distance                              & $d$           & $150\,\mathrm{pc}$\\
\hline
\end{tabular}
\tablefoot{
For more details on the parameter definitions, see \citet{Woitke2009a,Woitke2011,Woitke2016}.
\tablefoottext{a}{$\chi^\mathrm{ISM}$ is given in units of the Draine field \citep{Draine1996b,Woitke2009a}.}
\tablefoottext{b}{We use distributed hollow spheres for the dust opacity calculations \citep{Min2005,Min2016}.} 
\tablefoottext{c}{The optical constants are from \citet{Dorschner1995a} and \citet{Zubko1996c}.}}
\end{table}
\subsection{Radiation thermochemical modelling}
\label{sec:rtcmodelling}
To model the CPD we used the radiation thermochemical disk code \prodimo \citep{Woitke2009a,Kamp2010,Woitke2016}.\ The \prodimo code consistently solves for the dust radiative transfer, gas thermal balance, and chemistry for a given static two-dimensional dust and gas density structure. The results of this are the local disk radiation field, dust and gas temperature structure, and chemical abundances. Furthermore, \prodimo provides modules to produce synthetic observables such as spectral lines \citep{Woitke2011}, spectral energy distributions (SED) \citep{Thi2011}, and images.

\begin{figure*}
\includegraphics[width=0.485\hsize]{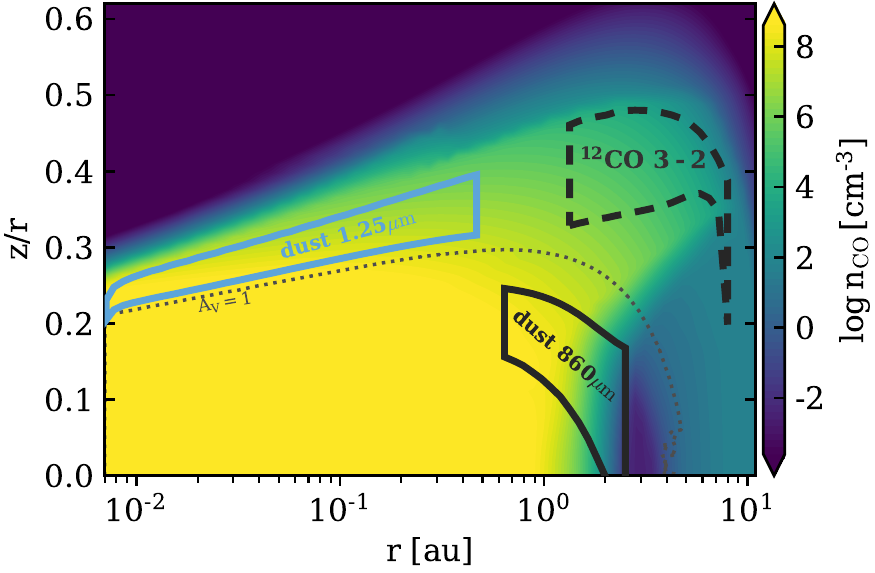}\hspace{0.2cm}
\includegraphics[width=0.485\hsize]{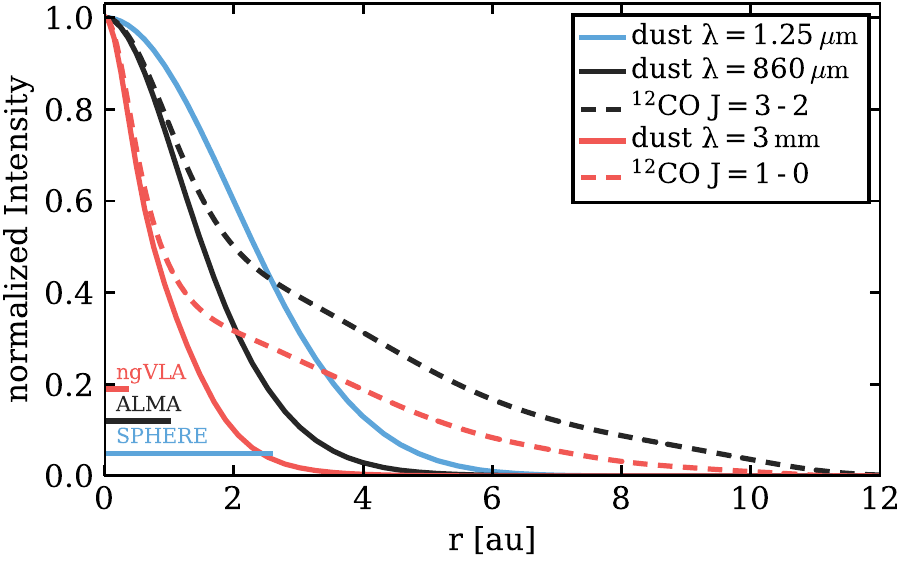}
\caption{\emph{Left panel: }Location of the main emitting region for the near-infrared and (sub)mm dust emission and the $^{12}\mathrm{CO}\,J\!=\!3\!-\!2$ line in the reference model. The vertical lines of the boxes denote the radii where the cumulative flux in radial directions reaches 15\% and 85\%, respectively. The horizontal lines denote the heights where the cumulative flux integrated vertically reaches 15\% and 85\% 
at each radial position. The coloured contours show the CO number density and the dotted grey line shows where the visual extinction $A_\mathrm{V}$ reaches unity. \emph{Right panel: }Normalized radial intensity profiles (azimuthally averaged) for the dust continuum and line images. Shown are profiles derived from images convolved with representative beam sizes for SPHERE (blue, $\mathrm{FWHM}=0.031\arcsec$), ALMA (black, $\mathrm{FWHM}=0.012\arcsec$), and ngVLA (red, $\mathrm{FWHM}=0.004\arcsec$).}
\label{fig:reference_observables}
\end{figure*}
For the chemistry, we used a chemical network including 86 chemical species and 1148 chemical reactions, including gas-phase chemistry, H$_2$ formation on grains, and ice chemistry (freeze-out; thermal, cosmic ray, and photo desorption). The chemical network is identical to the so-called small network of \citet{Kamp2017} except for the X-ray chemistry, which we did not include as we do not expect strong X-ray radiation from the planet. The gas-phase chemical reactions are based on the UMIST 2012 database \citep{McElroy2013b}. As shown in \citet{Kamp2017}, the CO abundance and resulting line fluxes are already stable for small networks and across different chemical databases. The chemical network used provides a sufficiently accurate treatment of CO chemistry and includes the main chemical heating/cooling agents and processes.

\subsection{Synthetic observables}
\label{sec:methodsyntheticobs}
We produced synthetic observables for the gas and dust in the (sub)mm regime, such as SEDs, spectral lines, and images at wavelengths ranging from $400-3000\,\mathrm{\mu m}$. This covers ALMA Bands~6 to 10 and Band~6 of the future ngVLA. This approach allowed us to determine the optimal wavelength/frequency for the observations of wide-orbit CPDs. For the gas we focussed on the spectral lines of $^{12}\mathrm{CO}$ as those lines are most likely the strongest emitters in the (sub)mm regime. For the line transfer calculations we used collisions rates for CO with $\mathrm{H_2}$ (Leiden Atomic and Molecular Database; \citealt{Schoeier2005c,Yang2010}), He \citep{Cecchi-Pestellini2002}, atomic hydrogen \citep{Balakrishnan2002}, and electrons \citep{Thi2013}.

To test if our models are actually observable, we compared them to the expected sensitivities for line and continuum observations in the various ALMA and ngVLA bands (see Appendix~\ref{sec:almasens}). We considered varying beam sizes and bandwidths for the synthetic observations but also present full ALMA/CASA simulations for a subset of models (see Appendix~\ref{sec:almasim} for details).

Most of the observed PMCs are located at a distance of $d\approx 150\,\mathrm{pc}$ (e.g. \citet{Wu2017}), hence we use that distance for all presented observables. For the inclination we take \mbox{$i=45^\circ$} except for our model of the CS~Cha companion  (Sect.~\ref{sec:cscha}), in which we use \mbox{$i=80^\circ$} as reported by \citet{Ginski2018}.
\section{Results}
\label{sec:results}
\subsection{Observational properties of the reference model}
\label{sec:res_reference}
First we discuss certain observational features of our reference model,
which are useful for the presentation and discussion of subsequent models. Fig.~\ref{fig:reference_observables} shows the main emitting regions of the near-infrared dust emission (VLT/SPHERE) as well as (sub)mm dust emission and $^{12}\mathrm{CO}$ line emission (ngVLA, ALMA). Furthermore we show a comparison of normalized radial intensity profiles to discuss the measured apparent disk radii. 

The $1.25\,\mathrm{\mu m}$ dust emission is concentrated to radii $r\lesssim1 \,\mathrm{au}$. For a SPHERE beam of $0.031\arcsec$ ($4.65\,\mathrm{au} $ at $d=150\,\mathrm{pc}$) \citep{Ginski2018}, the disk is therefore unresolved at distances of about $150\,\mathrm{pc}$. Although the real disk size of the reference model is $r_\mathrm{out}\approx11\,\mathrm{au}$ it appears as small as $r_\mathrm{out}\approx2\,\mathrm{au}$ in VLT/SPHERE images. This is consistent with the observation of \mbox{CS Cha c} where the potential CPD remains unresolved with VTL/SPHERE.

\begin{figure}
\centering
\resizebox{\hsize}{!}{\includegraphics{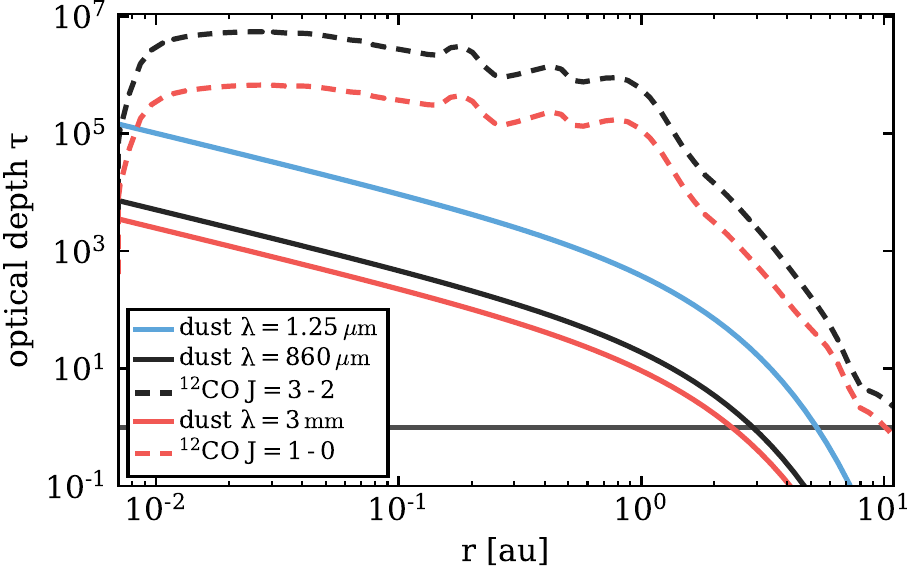}}
\caption{Vertical optical depth $\tau$ for the dust and for $^{12}\mathrm{CO}$ lines as a function of radius in the reference model (see also Fig.~\ref{fig:reference_observables}). The horizontal dark grey line indicates $\tau=1$.}
\label{fig:reference_taus}
\end{figure}

Compared to the near-infrared emission the $860\,\mathrm{\mu m}$ emission appears more compact with a half width half maximum of \mbox{$r\approx 1.7\,\mathrm{au}$}. This is due to the smaller ALMA beam of $0.012\arcsec\,(1.8 \,\mathrm{au}$ at $d=150\,\mathrm{au}$). However, compared to the \mbox{near-infrared} emission the main emission region is actually at larger radii (left panel of Fig.~\ref{fig:reference_observables}). We note that the $860\,\mathrm{\mu m}$ emission is optically thick out to $r\approx3\,\mathrm{au}$ (see Fig.~\ref{fig:reference_taus}). This indicates that the (sub)mm emission is dominated by the optically thick part of the CPD. 

The continuum non-detections of wide-orbit CPDs with ALMA indicate low dust masses or very compact optically thick disks ($r_\mathrm{out}\lesssim0.5\,\mathrm{au}$, \citealt{Wu2017}). Compared to those non-detections, the $860\,\mathrm{\mu m}$ disk in our reference model seems to be too large. We discuss this further in Sections~\ref{sec:results_struc} and \ref{sec:results_dust}.

In contrast to the dust emission the \mbox{$^{12}\mathrm{CO}\,J\!=\!3\!-\!2$} emission traces mostly the outer disk and can, in principle, be spatially resolved with ALMA even with larger beams. In the reference model the line is optically thick throughout the disk (see~Fig.~\ref{fig:reference_taus}) and traces only the upper layers of the disk. Because of the high optical depths \mbox{$^{12}\mathrm{CO}\,J\!=\!3\!-\!2$} is very sensitive to the disk size (i.e. emitting area) and the temperature. The different apparent extents of the dust and line emission in our model is similar to observations of T~Tauri disks \citep[e.g.][]{deGregorio-Monsalvo2013a,Ansdell2018}. It is still unclear if this difference is only caused by optical depth effects or if radial dust migration also plays a role \citep{Woitke2016,Facchini2017}. In our models the cause is solely the different dust and line optical depths.

For ngVLA observations at $3\,\mathrm{mm}$ the situation is similar to the ALMA Band~7 at $860\,\mathrm{\mu m}$. The main difference is the possible higher spatial resolution of the ngVLA ($0.004\arcsec, 0.6 \,\mathrm{au}$ at \mbox{$d=150\,\mathrm{au}$}), which might allow us to also resolve the dust disk (see right panel of Fig.~\ref{fig:reference_observables}). Compared to the $860\,\mathrm{\mu m}$ emission, the emission at $3\,\mathrm{mm}$ is weaker and the optically thick region of the dust and the gas are slightly smaller (see Fig.~\ref{fig:reference_taus}).
\subsection{Exploring disk mass and size}
\label{sec:results_struc}
\begin{figure}
\includegraphics[width=\hsize]{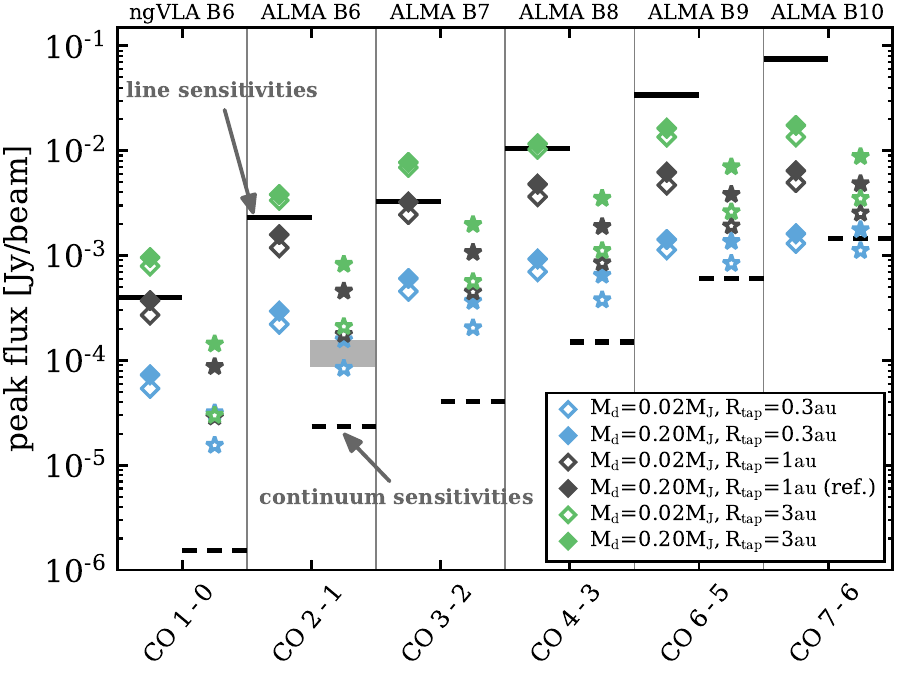}
\caption{Comparison of $^{12}\mathrm{CO}$ line and continuum peak fluxes from CPD models with varying structure properties (radius and mass) to $5\sigma$ detection limits for one band of ngVLA and various ALMA bands (see Appendix~\ref{sec:almasens} for details). The dust emission is reported at the same wavelength as the corresponding line. A beam of $0.1\arcsec\times0.1\arcsec$ is assumed. For the lines a channel width of $1\,\mathrm{km/s}$ was used. The peak flux is the maximum flux value in the synthetic images; for the lines all channels are considered to evaluate the peak flux (see Appendix~\ref{sec:almasim} for details). The diamond symbols show the fluxes for the lines and the star symbols the corresponding continuum fluxes. The black solid and dashed lines indicate the sensitivity limits for the line and continuum, respectively.  The light grey box shows the range of $3\sigma$ sensitivity levels as reported by \citet{Wu2017} for their ALMA Band 6 continuum non-detections of a sample of wide-orbit PMCs.}
\label{fig:cstruc}
\end{figure}

In Fig.~\ref{fig:cstruc} we compare the modelled fluxes for the dust continuum and gas lines to the $5\sigma$ detection limits of ngVLA and ALMA ($6\,\mathrm{h}$ on-source observing time, see Appendix~\ref{sec:almasens}). We report the peak(maximum) fluxes determined from the synthetic beam convolved continuum images and line cubes (see Appendix~\ref{sec:almasim}). If the peak flux of the model is above the reported sensitivity limit a $5\sigma$ detection is possible in the case of the lines in at least one channel. Fig.~\ref{fig:cstruc} shows the reference model and models with varying $R_\mathrm{tap}$ (i.e. changing the radial extent of the disk) and with a factor of 10 lower disk mass. Indicated in Fig.~\ref{fig:cstruc} are also the \mbox{$3\sigma$ rms} values reported by \citet[][Table 1]{Wu2017} for their observations of five potential CPDs with ALMA in Band 6. The average on-source integration time for their sample is around 12 minutes. 

As Fig.~\ref{fig:cstruc} shows our reference model is easily detectable in the dust, also with significantly shorter observing times (i.e. the flux is above the \citet{Wu2017} upper limits). For the lines only the \mbox{$^{12}\mathrm{CO}\,J\!=\!1-0$} (ngVLA Band 6) and \mbox{$^{12}\mathrm{CO}\,J\!=\!3-2$} (ALMA Band 7) can be detected with a S/N ratio of five. For the assumed  dust-to-gas mass ratio of $d/g=0.01$ the CO lines are therefore harder to detect than the dust. This is expected as for line detections much narrower bandwidths are required, compared to the continuum. The small disk models ($r_\mathrm{out}\approx4\,\mathrm{au}$) are roughly consistent with the upper limits of \citet{Wu2017} but are not detectable in the lines. For all models the best S/N is reached with the ngVLA and ALMA Band 7, whereas the ALMA high-frequency bands (Bands 8 to 10) are not suited for detections of CPDs such as those modelled in this work.

The change in the peak fluxes of the lines is mainly caused by the different disk sizes. This is a consequence of the high optical depth of the $^{12}\mathrm{CO}$ lines (see Fig.~\ref{fig:reference_taus}). The disk mass itself has no strong effect on the line emission. Actually, the lower fluxes of the low-mass models are mainly a consequence of smaller disk outer radii. Lowering the mass in our model makes the disk slightly smaller because of the way we define the disk outer radius (see also Fig.~\ref{fig:cStrucNH}). If the disk outer radii were adapted to the values of the higher mass models (e.g. by slightly increasing $R_\mathrm{tap}$) the fluxes would again become nearly identical.

Compared to the lines the dust emission is more affected by lowering the disk mass. In the model with ten times lower disk mass the \mbox{$\mathrm{^{12}CO}\,J\!=\!3\!-\!2$} line drops only by a factor of $1.3,$ whereas the corresponding dust flux drops by a factor of $2.4$ compared to the reference model. As seen in Fig.~\ref{fig:reference_taus} the line emission is significantly more optically thick than the dust and therefore the dust emission is more affected by lowering the disk mass. This can also be inferred from the slight decrease of this effect at shorter wavelengths, as the dust is more optically thick at short wavelengths. For \mbox{$^{12}\mathrm{CO}\,J\!=\!1\!-\!0$} (3 mm) the line drops by a factor of 1.4, if the disk mass is lowered by a factor of ten, whereas the dust emission drops by a factor of 3.
 
In contrast to our reference model our low-mass and compact disk models are consistent with the observational upper limits of \citet{Wu2017}. But those models can be detected with the much higher integration time of $6\,\mathrm{h}$. For example the model with $M_\mathrm{d}=0.02~M_\mathrm{J}$ and an outer disk radius of about $r_\mathrm{out}\approx9\,\mathrm{au}$ ($R_\mathrm{tap}=1\,\mathrm{au}$) might not have been detected by \citet{Wu2017} but can be detected in both the gas (only about $3\sigma$) and dust with the observing parameters assumed in this work. In those models the $860\,\mathrm{\mu m}$ continuum emission is optically thick out to $r\approx1.5\,\mathrm{au}$. This outer radius of the optically thick (sub)mm disk is about a factor of three larger than the upper limit of $r_\mathrm{out}\lesssim0.5\,\mathrm{au}$ derived by \citet{Wu2017} for their sample, but is smaller than the upper limit of $r_\mathrm{out}\lesssim2.9\,\mathrm{au}$ derived by \citet{Wolff2017} for \object{DH Tau b}. However, qualitatively speaking our results from the full radiative-transfer models are consistent with the analytical approach of \citet{Wu2017} and indicate that the radial extent of the dust emission is small in CPDs. In case of $d/g=0.01$ the gas lines are likely not detectable for such small disks ($r_\mathrm{out}\lesssim9\,\mathrm{au}$). 

To summarize, our simulations indicate that CO spectral lines can be detected in CPDs if the gas disks are as large as $r_\mathrm{out}\approx10\,\mathrm{au}$ at target distances of $d=150\,\mathrm{pc}$. The ALMA Band~7 is best suited for line detections, but also the ngVLA $3\,\mathrm{mm}$ Band might work as well. For the higher frequency bands a detection becomes unlikely.
\subsection{Dust properties and evolution}
\label{sec:results_dust}
Similar to protoplanetary disks around solar-like stars, CPDs likely experience significant dust evolution. Especially interesting in this context is radial inward migration of large grains and consequently a depletion of the dust disk. Studies of \mbox{brown-dwarf} disks \citep{Pinilla2013} and analytical estimates of \citet{Zhu2018} have suggested that the radial migration process is much more efficient in CPDs (or compact disks in general) than for disks around T~Tauri stars. This is mainly due to the lower central mass/luminosity and disk surface density (see e.g. \citealt{Zhu2018} Eq. 17). As the expected gas accretion timescales are significantly longer \citep{Zhu2018}, it is therefore possible that at least some of the currently known CPD candidates are dust poor but gas rich as they are still accreting.

We tested the rapid dust evolution scenario with the dust evolution code \mbox{\emph{two-pop-py}} \citep{Birnstiel2012} for our our reference model (see Appendix~\ref{sec:dustevol} for details and a comparison to a T~Tauri disk model). We find that the dust rapidly migrates inward and already after about $10000\,\mathrm{yr}$ the dust-to-gas mass ratio drops to $d/g=10^{-3}$ in most regions of the disk and reaches values of $\lesssim10^{-4}$ at $1~\mathrm{Myr}$. Those results are consistent with the analytical estimates of \citet{Zhu2018}. Not included in this model are possible dust traps due to pressure bumps \citep[e.g.][]{Pinilla2013}. \citet{Draedilzkowska2018} investigated such a scenario for CPDs by means of hydrodynamical modelling. They found that dust can be indeed trapped very close to the planet at about $r=0.05\,\mathrm{au}$. Trapping of dust so close to the planet would leave most of the CPD dust depleted and results in a very small but heavily optically thick dust disk.

\begin{figure}
\includegraphics[width=\hsize]{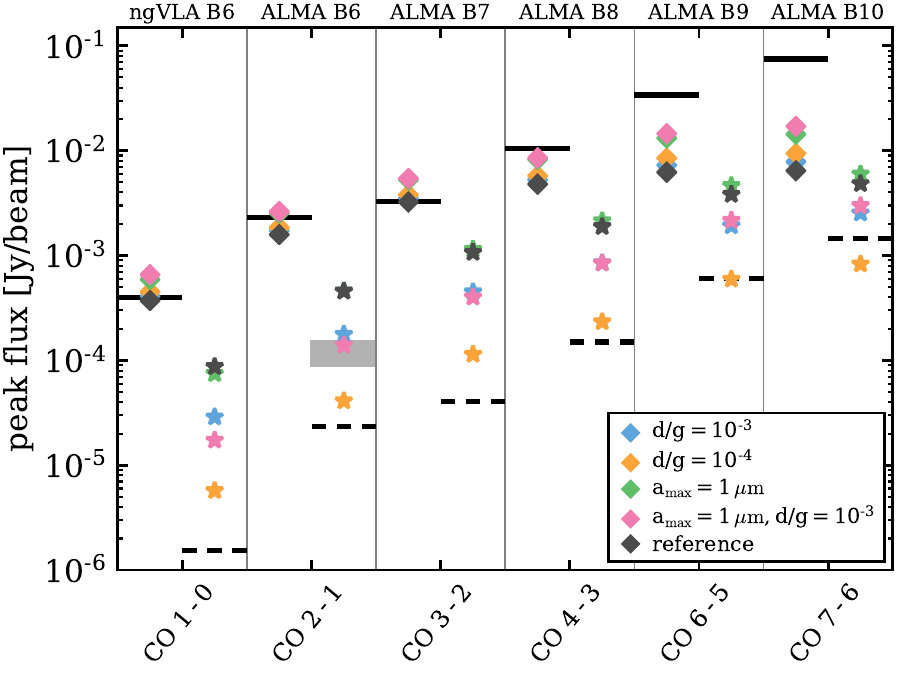}
\caption{Same as Fig.~\ref{fig:cstruc} but for models with varying dust properties. The underlying model is always the reference model with $L_\mathrm{p}=10^{-2}\,L_\mathrm{\sun}$, $M_\mathrm{d}=0.2\,M_\mathrm{J}$, and $r_\mathrm{out}\approx11\,\mathrm{au}$.}
\label{fig:cDust}
\end{figure}
To study the impact of dust evolution on the observational properties of CPDs we present models with varying dust properties such as maximum grain size and/or the total dust-to-gas mass ratio. We compare those models with the reference model in Fig.~\ref{fig:cDust}. Only the listed dust properties have changed, all other parameters are identical to the reference model. From Fig.~\ref{fig:cDust} we see that changing the dust disk has only marginal impact on the CO line emission. In the models with a maximum dust grain size of $a_\mathrm{max}=1\,\mathrm{\mu m}$ the disk becomes warmer and the line fluxes, in particular for the higher J lines increase. Decreasing $d/g$ by removing dust mass has even less impact on the line fluxes.

For the dust emission changes are more pronounced. Interestingly for the \mbox{$a_\mathrm{max}=1\,\mathrm{\mu m}$} model the (sub)mm dust emission is not significantly affected although the emission at those wavelengths should be dominated by large dust grains \mbox{($a_\mathrm{grain}\gtrsim100\,\mathrm{\mu m}$)}. In the models presented in this work, the (sub)mm dust emission is dominated by optically thick emission and the sensitivity to the dust grain size is therefore lost. In the $a_\mathrm{max}=1\,\mathrm{\mu m}$ model, the $860\,\mathrm{\mu m}$ emission is optically thick up to $r\approx1.7\,\mathrm{au}$ and the average dust temperature in the emitting region is $\approx34\,\mathrm{K}$ compared to $2.9\,\mathrm{au}$ and $\approx27\,\mathrm{K}$ for the reference model. So the higher temperature compensates for the smaller emitting area in those models.

In the models with reduced dust-to-gas mass ratios, the fluxes for the dust emission drop by factors of three for $d/g=10^{-3}$ and by more than an order of magnitude for $d/g=10^{-4}$. We reduced the dust mass homogeneously over the whole disk, therefore the optically thick dust disk becomes smaller and the emission decreases. In the model with $d/g=10^{-3}$ the $860\,\mathrm{\mu m}$ emission is still optically thick up to $r\approx1.3\,\mathrm{au}$. In the $d/g=10^{-4}$ the emission starts to be dominated by optically thin emission and the disk is only optically thick up to $r\approx0.4\,\mathrm{au}$. Such dust depleted disks are consistent with the detection limits of \citet{Wu2017}.

Strong dust depletion in CPDs is an alternative scenario for non-detections of CPDs in the (sub)mm continuum around wide-orbit PMCs. In that case the dust disks also appear small at (sub)mm wavelengths. The main difference to the small disk scenario is that the line emission is mostly unaffected in the dust depletion scenario simply because the gas disk structure is unaffected. In the dust depletion scenario, the gas lines are still detectable but the continuum emission is weak and the dust disk appears significantly smaller than the gas disk. 
\begin{figure}
\includegraphics[width=\hsize]{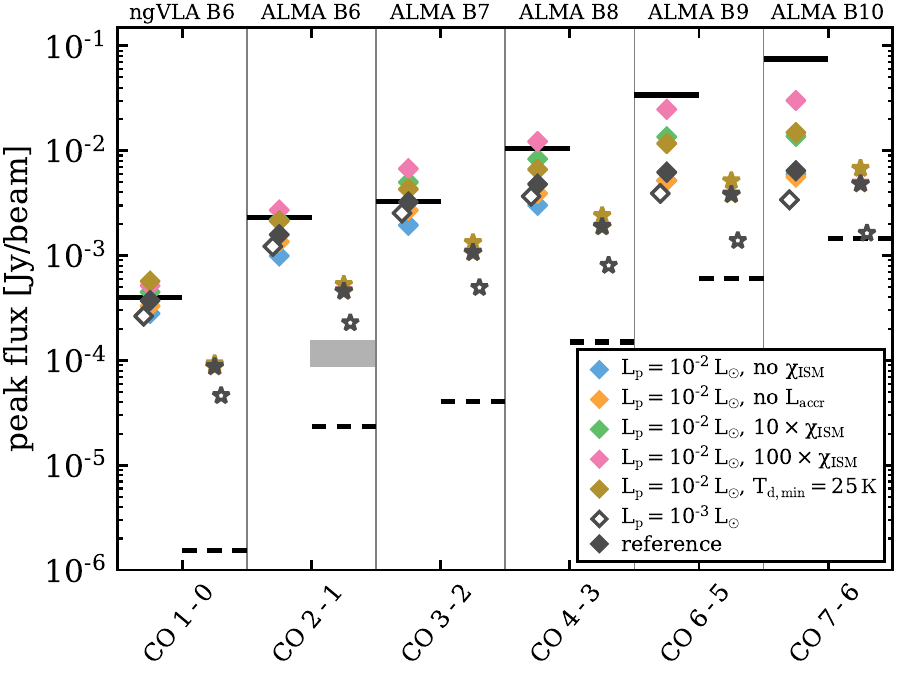}
\caption{Same as Fig.~\ref{fig:cstruc} but for models with varying radiation properties of the PMC and the environment (background field). The underlying disk structure is always that from the  reference model.}
\label{fig:cRad}
\end{figure}
\subsection{Radiative environment}
\label{sec:results_rad}
In this section we investigate the importance of the various radiation sources such as planetary luminosity, accretion luminosity, and background radiation field for the line and dust emission. The density structure of those models is identical to the reference model. In Fig.~\ref{fig:cRad}, we show the fluxes of those models in comparison to the expected telescope sensitivities.

For the model with $L_\mathrm{p}=10^{-3}\,L_\mathrm{\sun}$ but same accretion luminosity as the reference model, the line fluxes drop by factors of 1.3 to 1.9, where $\mathrm{CO}\,J=3-2$ is the least and $\mathrm{CO}\,J=6-5$ the most affected line. The continuum fluxes drop by factors of 1.9 to 3 compared to the reference model, with the lowest change in the ngVLA band and the highest change in ALMA B10. The dust temperature is more sensitive to the optical emission of the companion and is therefore more affected by the decrease of the photospheric luminosity. For the gas, the far-ultraviolet (far-UV) radiation is more important and as the accretion luminosity nor the interstellar medium (ISM) background field changed in this model, the gas temperature and line fluxes are not significantly affected. However, even in the low luminosity model, the dust emission in ALMA B6 is still higher than the sensitivity limits of \citet{Wu2017}.

For the model without an ISM background field (no $\chi$) the line emission drops by factors of 1.1 (ALMA B10) to 1.7 (ALMA B7), whereas in the model without accretion luminosity (no $L_\mathrm{accr}$) the lines drop at most by a factor of about $1.2$ (similar in all bands). In both cases the dust is not affected as the dust temperature is not sensitive to the far-UV radiation. This shows that the background radiation is actually more important for the lines than the accretion luminosity, especially at long wavelengths. The reason is that radiation of the PMC only impacts the disk gas temperature for radii $r\lesssim1\,\mathrm{au}$ (see also Appendix~\ref{sec:reftemp}), whereas the ISM background field has an impact on the entire disk. In the reference model, the ISM field is dominant for $r\gtrsim4\,\mathrm{au}$, with respect to the stellar UV. For similar models but with  $L_\mathrm{p}=10^{-3}\,L_\mathrm{\sun}$ (not shown) the relative importance of the far-UV radiation slightly increases, but otherwise the situation is similar. These results show that the presence of the interstellar background radiation field makes the line emission insensitive to the radiation properties of the PMC. However, we note that for very compact disks, the radiation of the PMC is more important.

In the case of CPDs the host star can also provide an additional ``background'' radiation field. The importance of this contribution depends on the orbit of the companion, presence of a disk around the stellar host (i.e. shielding of the CPD), and size of the CPD itself. For example \citet{Ginski2018} concluded that for their compact \object{CS~Cha} CPD model ($r_\mathrm{out}=2\,\mathrm{au}$) the contribution from the stellar host is negligible. However, \cite{Wolff2017} found for their model of the DH~Tau system with a large CPD ($r_\mathrm{out}\approx70\,\mathrm{au}$) that the stellar radiation can efficiently heat the outer region of the CPD to temperatures of $22\,\mathrm{K}$. 

To study the impact of enhanced background radiation, we show in Fig.~\ref{fig:cRad} models with a 10 and 100 times stronger ISM radiation field ($\chi_\mathrm{ISM}=10,~\chi_\mathrm{ISM}=100$) and a model where we assume that the additional stellar radiation heats the dust disk to a minimum temperature of $T_\mathrm{d,min}=25\,\mathrm{K}$. Relative to the reference model the line fluxes increase in all three cases whereas the dust is only affected in the model with $T_\mathrm{d,min}=25\,\mathrm{K}$. We note that for the $\chi_\mathrm{ISM}=100$ model CO is efficiently photodissociated in the outer disk and the $^{12}\mathrm{CO}\,J\!=\!3\!-\!2$ emission becomes optically thin at $r\approx9\,\mathrm{au}$ compared to $r\approx10\,\mathrm{au}$ in the reference model. However, the higher gas temperatures easily compensate for the smaller emitting region.

Generally speaking, the size of the CPD has more impact on the strength of the line emission than the luminosity of the PMC. The gas temperature is less sensitive to the PMC emission than the dust temperature because the ISM \mbox{far-UV} background field compensates for the lack of PMC emission. The presence of an enhanced background field, from the ISM or the star, only boosts the line emission and makes the detection of lines from CPDs easier compared to our reference model.
\begin{figure} 
\includegraphics[width=\hsize]{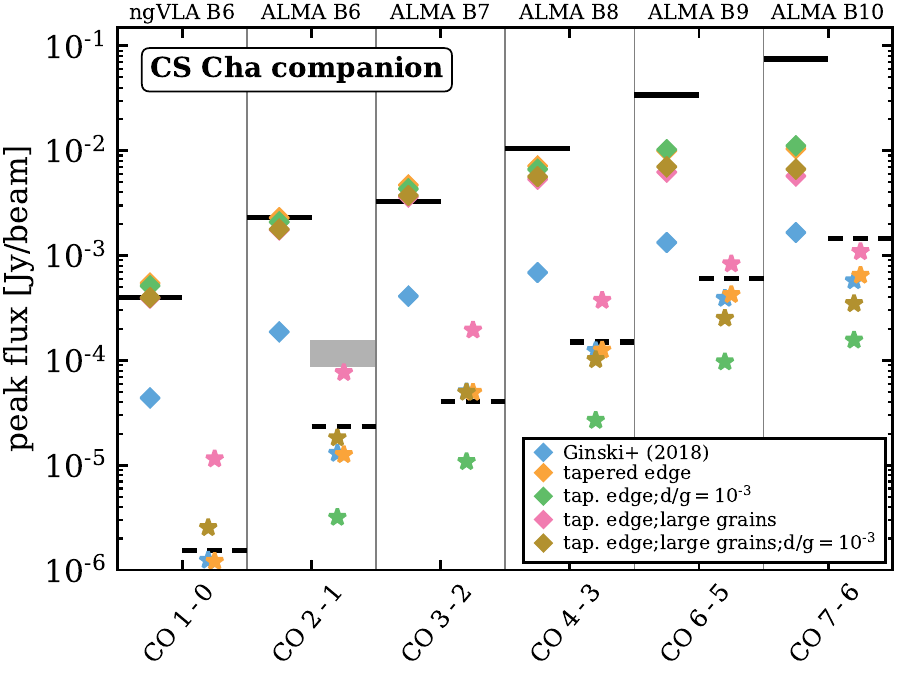}
\caption{Same as Fig.~\ref{fig:cstruc} but for the CS Cha companion model. Shown are the original \citet{Ginski2018} model and several tapered-edge models with  different dust-to-gas mass ratios ($d/g$) and dust compositions/populations (large grains).}
\label{fig:cschasens}
\end{figure}
\begin{figure*}
\includegraphics[width=\hsize]{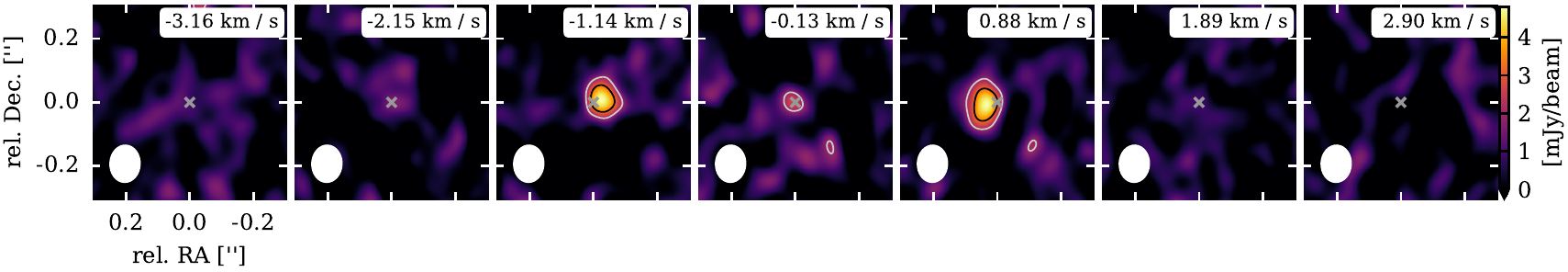}
\caption{$^{12}\mathrm{CO}\,J\!=\!3\!-\!2$ (ALMA B7) synthetic channel intensity maps for a CS~Cha companion model. The grey cross indicates the location of the PMC. The white ellipse in each panel shows the beam with a size of $0.11\arcsec\times0.09\arcsec$. In the top right corner of each panel the  velocity relative to the systemic velocity is indicated. The grey and black contours show 3 and 5 times the theoretical rms level of $0.65~\mathrm{mJy/beam}$ (see Table~\ref{table:sensitivity}). The measured rms levels in the shown channels are approximately in the range $0.49-0.76\,\mathrm{mJy/beam}$. The measured peak flux is about $4.8\,\mathrm{mJy/beam}$.}
\label{fig:cschaalma}
\end{figure*}
\subsection{CS Cha companion}
\label{sec:cscha}
In \citet{Ginski2018} dust radiative transfer models for the circumplanetary material of the CS~Cha companion are presented, including a ``disk only'' model that matches the photometric data ($0.5-3.5\,\mathrm{\mu m}$) and the polarization measurements. This model features a PMC with $M_\mathrm{p}=20\,M_\mathrm{J}$, $T_\mathrm{p}=2500\,\mathrm{K}$, $L_\mathrm{p}\approx2.5\times10^{-3}\,L_\mathrm{\sun}$ and a highly inclined CPD ($i=80^\circ$) with a dust mass of $2\times10^{-4}\,M_\mathrm{J}$ (10 times lower than in our reference model). For the dust opacities they use astronomical silicates \citep{Weingartner2001c} with a single grain size of $1\,\mathrm{\mu m}$. As \mbox{\object{CS Cha c}} is unresolved with VLT/SPHERE, they fixed the outer disk radius to $R_\mathrm{out}=2\,\mathrm{au}$. The goal of the modelling in \citet{Ginski2018} was not to find the best-fitting model, but to show that a CPD can explain the main features of their observations.

We use the \mbox{CS Cha c} dust disk model of \citet{Ginski2018} as a starting point for our gas disk modelling to see if the potential CPD is detectable in the (sub)mm regime. The unresolved optical and near-infrared data provides only limited constraints on the dust properties (i.e. grain size) and on the gas disk. We therefore also constructed models in a similar fashion as our models presented in Sections \ref{sec:results_struc} and \ref{sec:results_dust} but always check whether they are still consistent with observational constrains.

As discussed in Sect.~\ref{sec:res_reference} and as shown in Fig.~\ref{fig:reference_observables} the scattered light images are not necessarily a good tracer for the real disk size. We therefore also constructed models with $r_\mathrm{out}\approx r_\mathrm{Hill}/3$ by using a tapered outer edge disk structure. Assuming a semimajor axis $a=215\,\mathrm{au}$ (projected separation) for the orbit, a combined mass of the central binary of $M_\mathrm{*}=1\,M_\mathrm{\sun}$ and $M_\mathrm{p}=20\,M_\mathrm{J}$ \citep{Ginski2018}, we obtain $r_\mathrm{Hill}/3=13.2\,\mathrm{au}$ for CS~Cha~C. To estimate $r_\mathrm{Hill}$ we used a distance of $d=165\,\mathrm{pc}$ to be consistent with \citet{Ginski2018}. We note that recent GAIA distance estimates place the Chameleon~I star formation region at $179\pm10\,\mathrm{pc}$ \citep[GAIA DR1, ][]{Voirin2018} or even at $192\pm6\,\mathrm{pc}$ \citep[GAIA DR2,  ][]{Dzib2018}, whereas the distance derived from the parallax reported in the GAIA DR2 catalogue \citep{GaiaCollaboration2016,GaiaCollaboration2018,Lindegren2018} of CS Cha is $d=176\pm1\,\mathrm{pc}$. Anyway, such distance variations do not have a significant impact on our results because for larger distances $r_\mathrm{Hill}$ would also increase and the resulting fluxes of the models would remain very similar (see also Fig.~\ref{fig:hillradii}).

In Fig.~\ref{fig:cschasens} we show the expected fluxes for a couple of CS~Cha CPD models. For the first model we simply used the parameters of \citet{Ginski2018}. All other models are \mbox{tapered-edge} models with $r_\mathrm{out}=r_\mathrm{Hill}/3$ (i.e. using $R_\mathrm{tap}=1.75\,\mathrm{au}$). For those, we also show models with $d/g=10^{-3}$ and models that use the dust properties of our reference model instead of the $1\,\mathrm{\mu m}$ grains used by \citet{Ginski2018}. It is apparent from Fig.~\ref{fig:cschasens} that the original model of \citet{Ginski2018} would not be detected in the lines and even a dust detection is unlikely in the (sub)mm regime. In addition to the small disk, the dust opacity used by \citet{Ginski2018} is about an order of magnitude lower in the (sub)mm compared to the dust opacity used for our modelling. The reason for this is that we consider larger grain sizes and include amorphous carbon for the dust grain  composition \citep[see e.g. ][]{Woitke2016}. As a consequence, the dust disk is optically thin at (sub)mm wavelengths in the Ginski model and the fluxes are significantly lower compared to our low-mass and small disk models shown in Fig.~\ref{fig:cstruc}. Additionally, the higher inclination for the \mbox{CS Cha c} model reduces the peak fluxes for the gas and dust compared to a model with $i=45\degr$ as the disk appears smaller (e.g. in ALMA B7 by factors of 1.4 and 1.7 for the gas and dust, respectively). The dust opacities, higher inclination, and lower total mass increase the gas-to-dust peak flux ratios in the \mbox{CS Cha c} models compared to our reference model.

In the model with a \mbox{tapered-edge} and $r_\mathrm{out}\approx r_\mathrm{Hill}/3$ (all other parameters are unchanged) the lines move into the observable regime as expected due to the much larger gas disk. However, the apparent radius of the dust disk at $1.25\,\mathrm{\mu m}$ of $5\,\mathrm{au}$ is inconsistent with observations, as the disk would have been resolved with VLT/SPHERE. Even if we lower the dust mass by a factor of ten ($d/g=10^{-3}$) or switch to the dust properties of our reference model (larger grains) the dust disk still appears too large. The dust disk at $1.25\,\mathrm{\mu m}$ is only small enough to be unresolved with VLT/SPHERE if we use $d/g=10^{-3}$ and our dust properties. Except for the apparent disk extension, all those models produce very similar photometric fluxes at micrometer wavelengths and are consistent with the SED presented in \citet{Ginski2018}.

This exercise shows that it is possible to construct disk models for \mbox{CS Cha c} with $r_\mathrm{out}\approx r_\mathrm{Hill}/3$ that are consistent with constraints from dust observations and are detectable in the lines. For the model with a tapered edge, large grains, and $d/g\!=\!10^{-3}$, we produced realistic ALMA simulations for $^{12}\mathrm{CO}\,J\!=\!3\!-\!2$ (see Appendix~\ref{sec:almasim} for details). The parameters of the simulations were chosen to give rms noise levels similar to those estimated with the sensitivity calculator (Appendix~\ref{sec:almasens}) and to have a similar beam size and channel width as used to estimate the peak fluxes shown in Fig.~\ref{fig:cschasens}. 

In Fig.~\ref{fig:cschaalma}, we show the seven central channels of the simulated $^{12}\mathrm{CO}\,J\!=\!3\!-\!2$ line cube. We see a clear detection ($>3\sigma$) in the three central channels. This is expected as the full width at half maximum (FWHM) of $^{12}\mathrm{CO}\,J\!=\!3\!-\!2$ line in the model is $3.14\,\mathrm{km\,s^{-1}}$ and the spectral resolution is only $1\,\mathrm{km\,s^{-1}}$. Despite the low spectral resolution, we can still see the signature of Keplerian disk rotation. The peak of the blue-shifted channel is just right of the centre (indicated by the grey cross) whereas the red-shifted channel peaks left of the centre. Although we used only a modest spatial and spectral resolution, the simulation shows that such observations can already trace the rotation pattern of the disk. 

With the low spectral and spatial resolution used, it is not possible to produce moment 1 maps or position-velocity diagrams that allow an accurate estimate of the central mass. Nevertheless, we produced full ALMA simulations for models with four times higher ($M_\mathrm{p}=80\,M_\mathrm{J}$) and four times lower ($M_\mathrm{p}=5\,M_\mathrm{J}$) PMC mass (see Appendix~\ref{sec:cschaalma_more} and Fig.~\ref{fig:cschaalma_more}). If we keep all other model properties the same (i.e. density structure and temperature structure) only the FWHM of the line profiles changes. In the high-mass case, the FWHM increases from $3.14$ to $5.76\,\mathrm{km\,s^{-1}}$, whereas for the low-mass PMC, the FWHM decreases to $1.76\,\mathrm{km\,s^{-1}}$. For the high-mass model, this means that now we can also see a signal in higher velocity channels (five channels in total), whereas in the low-mass case, we now only get a clear signal in the central channel. If the outer radius is also adapted according to the PMC mass to $r_\mathrm{out}\approx r_\mathrm{Hill}/3$, a detection for the low-mass model becomes unlikely, but for the \mbox{high-mass} model the disk and its velocity profile is easily detectable. However, those examples indicate that even for modest spectral and spatial resolutions it is feasible to constraint the PMC mass within a factor of a few.

Fig.~\ref{fig:cschaalma} demonstrates that it is possible to detect the potential CPD in the CS~Cha system with ALMA in about $6\,\mathrm{h}$ on-source time, if the gas disk is as large as $r_\mathrm{Hill}/3$. We note that for shorter observing times of about $3 \mathrm{h}$, we would still get a $3\sigma$ signal for the $^{12}\mathrm{CO}\,J\!=\!3\!-\!2$ line. It is feasible to constrain the mass of the CS~Cha companion within a factor of a few (e.g. exclude the case of a low-mass stellar companion) even with the low spectral and spatial resolution required to detect the CPD in the first place.
\section{Discussion}
\label{sec:discussion}
\subsection{Observing wide-orbit circumplanetary disks}
Our models show that a detection of CPDs around wide-orbit PMCs is doable with ALMA within 3 to 6 h of on-source integration time. Such observations can even provide a clear signature of a Keplerian rotation profile. However, certain conditions have to be met. 

\begin{figure}
\includegraphics[width=\hsize]{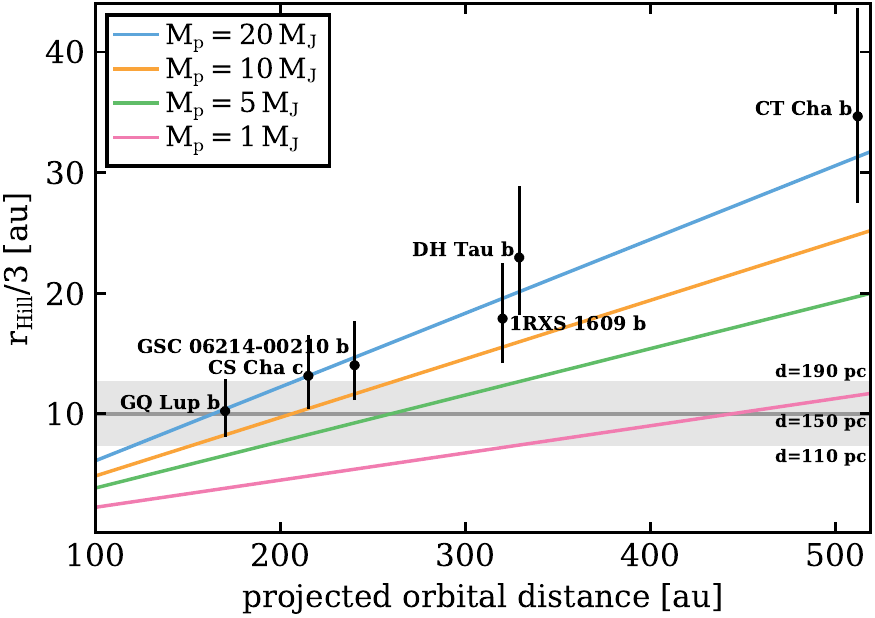}
\caption{One-third of the Hill radius ($r_\mathrm{Hill}/3$) vs. orbital distance for a selection of detected PMCs (black symbols) and as reference for companions orbiting a solar-mass star (coloured lines, 4 different PMC masses). We note that for example the curve for $M_\mathrm{p}=20\,M_\mathrm{J}$ is identical to the curve for a $M_\mathrm{p}=10\,M_\mathrm{J}$ companion but around a $0.5\,M_\mathrm{\sun}$ star. The thick grey horizontal line and the grey box indicate the detection limit for the gas disk of $r=10\pm2.7\,\mathrm{au}$ for distances of $110-190\,\mathrm{pc}$. The error bars indicate a factor of two uncertainty in the mass of the companion; the real error might be significantly larger.}
\label{fig:hillradii} 
\end{figure}

As we have shown the size of the CPD is the main criterion for a detection of the gas component. This argument holds as long as the CPD disk mass is high enough so that the $^{12}\mathrm{CO}$ lines remain optically thick throughout the disk. Under this assumption and neglecting properties such as the PMC luminosity and detailed disk structure, it is possible to relate the detectability of a gaseous CPD solely to its Hill radius. This allows us in a simple way to determine for which of the known wide-orbit PMC candidates a detection of their potential CPD is feasible with (sub)mm telescopes. In Fig.~\ref{fig:hillradii}, we show $r_\mathrm{Hill}/3$ as a function of orbital distance for a selection of wide-orbit CPD candidates (all show signs of ongoing accretion) and for companions with different masses orbiting a one solar-mass star. The data to calculate the Hill radius of the CPD candidates were collected from the literature \citep{Schmidt2008,Bowler2011,Wu2015,Wu2015a,Wu2017,Wolff2017,Ginski2018,Pearce2018}. For \mbox{CT Cha} and \mbox{GSC 06214-00210} we used the new GAIA DR2 distances of $d=192\,\mathrm{pc}$ \citep{Dzib2018} and $d=109\,\mathrm{pc}$ \citep{Pearce2018}, respectively. For all other targets the distances are $d\approx140-160\,\mathrm{pc}$.

\begin{figure*}
\includegraphics[width=\hsize]{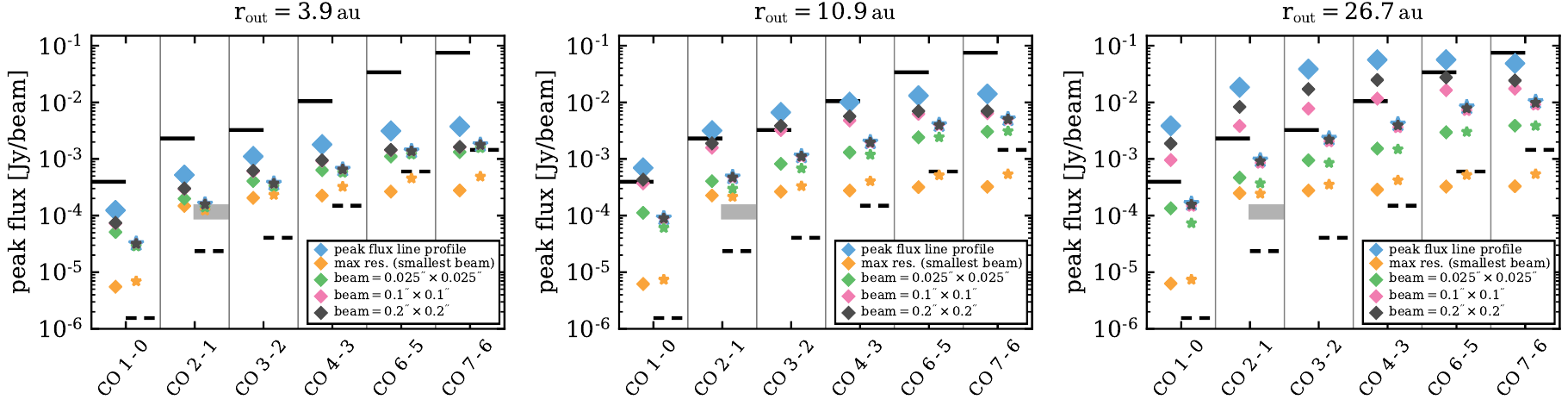}
\caption{Impact of the beam size on the observed fluxes. Each individual panel shows the same kind of plot as shown in Fig.~\ref{fig:cstruc}. In each panel one model with a certain disk extension is shown. For each of these models, the peak fluxes using different beam sizes are reported (coloured symbols in each panel). As a reference also the given line peak flux (blue diamonds) as derived from the line profile is shown (i.e. is not per beam). For the continuum, the big blue stars show the total integrated flux.}
\label{fig:beams}
\end{figure*}
According to our modelling results, CPDs need to have an outer radius of $r_\mathrm{out}\gtrsim10\,\mathrm{au}$ to be detectable at distances of about $d\approx150\,\mathrm{pc}$ (see e.g. Sect.~\ref{sec:results_struc}). More precisely the $^{12}\mathrm{CO}$ emission needs to be optically thick throughout the disk. For the \mbox{$^{12}\mathrm{CO}\,J\!=\!3\!-\!2$} line, this is true for disk masses as low as \mbox{$M_\mathrm{d}\approx10^{-4}\,M_\mathrm{J}$} assuming a radial surface density profile $\Sigma\propto r^{-1}$. We verified that this low-mass CPD is still detectable assuming the same companion parameters as for our reference model. As Fig.~\ref{fig:hillradii} shows, even for a CPD around a $M_\mathrm{p}=5\,M_\mathrm{J}$ companion orbiting a solar-mass star, detections are possible if $a\gtrsim250\,\mathrm{au}$ or $a\gtrsim200\,\mathrm{au}$ for $M_\mathrm{*}=0.5\,M_\mathrm{\sun}$. Fig.~\ref{fig:hillradii} also shows that for all considered CPD candidates their theoretically expected disk size of $r_\mathrm{Hill}/3$ is above the disk size limit we derived and therefore a detection of their gas disk is feasible. We want to emphasize that the errors on the companion mass estimates can be significant as the masses are only constrained by optical and near-infrared photometry (see e.g. \citealt{Wu2017a} for FW~Tau).

In addition to the disk size, the age of the system is another important factor to consider. It is reasonable to assume that in older systems the dissipation process for both the protoplanetary and the CPD are already advanced compared to younger objects, leading to less massive and smaller disks. For \object{1RXS~1609~b} an age of $10\,\mathrm{Myr}$ \citep{Wu2017} is suggested and for \object{GSC~6214-210~b} the age was recently revised to $17\,\mathrm{Myr}$ \citep{Pearce2018}. Hence they might not be the ideal candidates for a detection experiment. On the other hand studying these old systems (all other objects are about $2\,\mathrm{Myr}$ old; \citealt{Wu2017}) can provide insight into the evolution of PMCs and their CPDs. Especially interesting in this case is \mbox{GSC~6214-210~b} as it has an accretion rate of $\dot{M}_\mathrm{accr}\approx1.2\times10^{-11}\,M_\mathrm{\sun}\mathrm{yr^{-1}}$ 
\citep{Bowler2011}, which is actually higher than for some of the younger targets, indicating that it might still have a detectable CPD.

The expected disk size is also relevant for the choice of the spatial resolution. In Fig.~\ref{fig:beams} we show the impact of the beam size on the measured line fluxes for three models with different disk sizes. This clearly shows that for large disks and large beams the required on-source integrations times to detect the CPDs would be significantly lower than $6~\mathrm{h}$. On the other hand, a too small beam can prevent a detection. In Sect.~\ref{sec:results}, we always used a beam size of 0.1\arcsec , which is nearly an ideal choice for disks with \mbox{$r_\mathrm{out}\approx10\,\mathrm{au}$} at distances of $d\approx150\,\mathrm{pc}$ and is still good enough to detect the rotation signature (see Fig.~\ref{fig:cschaalma}). As the CPD size is not known a~priori the expected Hill radius can be used as a guide to select the optimal beam size for the observations. 

To detect a possible Keplerian signature, a high enough spectral resolution is required.  As we have shown in Sect.~\ref{sec:cscha} about $1\,\mathrm{km\,s^{-1}}$ bandwidth is good enough to spectrally resolve the line for a potential CPD around \mbox{CS Cha c}. The main factors influencing the  FWHM of the spectral line are the PMC mass (see Sect.~\ref{sec:cscha}) but also the disk size. For example in the models with three different disk sizes shown in Fig.~\ref{fig:cstruc} the FWHM for the $^{12}\mathrm{CO}\,J\!=\!3-2$ are 4.4, 2.5 and $1.7\,\mathrm{km\;s^{-1}}$ (from small to large disks). Considering such factors also allows us to optimize the observing strategy. For example the spectral resolution for targets on wider orbits should be higher because their disks might be larger.

This discussion shows that a number of factors need to be considered to derive the best observing strategy for the detection of a gas disk around a wide-orbit PMC. The flexible model presented in this work allows us to make accurate predictions for both the dust and gas observables of the potential CPD candidates. As observing such CPDs will be time consuming even with ALMA or ngVLA, such models are crucial to optimize observing programmes depending on the known properties of the companion and its host star.
\subsection{What can we learn from gas observations}
\label{sec:whatlearn}
\subsubsection{Companion and disk properties}
Current continuum observations of wide-orbit PMCs indicate small dust disks, either because they are unresolved \citep{Ginski2018} or not detected at all \citep{Wu2017}. However, in a compact CPD the dust might evolve on much shorter timescales than in \mbox{T Tauri disks} and rapid radial migration can lead to strongly depleted dust disks (see Appendix~\ref{sec:dustevol}). This implies that the gas disk might live longer than the dust disk, which increases the chances for a detection with respect to the continuum (see e.g. Sect.~\ref{sec:cscha}). Furthermore, the uncertainty due to the possible rapid dust evolution makes total disk mass estimates derived from continuum observations more unreliable as the assumption of a total gas-to-dust mass ratio of $100$ is not well justified. With the kind of observations proposed in this work, it is possible to derive useful constraints on the disk gas mass, even in case of non-detections. This would provide first constraints on the gas-to-dust mass ratio.

The CO gas observations are a better tracer of disk size than continuum observations, as in the (sub)mm the dust is less optically thick than the $^{12}\mathrm{CO}$ line. Assuming that the temperature structure of the CPD is known reasonably well, this is even possible if the gas disk is not spatially resolved, as the flux of optically thick line emission scales with the size of the emitting area (see also \citet{Greenwood2017} for brown-dwarf disks).

Another advantage of spectral line observations is that they also provide information on the velocity structure of the circumplanetary material. The hydrodynamic simulations of \citet{Szulagyi2016} suggested that giant gas planets do not form a CPD but an envelope-like structure. If such an object is scattered to a wide orbit, it likely evolves towards a Keplerian disk, but depending on the timescale, they might still show \mbox{non-Keplerian} velocity signatures. 

In case of the detection of a Keplerian profile it is also possible to constrain the mass of the PMC within a factor of a few, even with modest or low spectral and spatial resolution (see Sect.~\ref{sec:cscha}). Such observations might therefore answer the question if wide-orbit PMCs are actually proper planets.
\subsubsection{Formation scenarios}
\label{sec:formationscenario}
So far the main formation mechanism of wide-orbit PMCs remains unclear. Proposed scenarios include the formation in a fragmenting protostellar disk, core accretion, and subsequent scattering to wide orbits and turbulent fragmentation of molecular clouds (similar to binary formation). However, it seems that none of these scenarios can explain all the properties of the currently known PMC population (see e.g. \citealt{Wolff2017, Vorobyov2013,Stamatellos2015,Bryan2016}).

Small and compact CPDs are favoured in the core+scattering scenario as it is unlikely that the CPD will fully survive the ejection out of the parent protoplanetary disk. However, the survey of seven wide-orbit PMCs of \citet{Bryan2016} does not show any indication of other bodies with masses $>7 M_\mathrm{J}$ that could act as the ``scatterer'', which makes the scattering scenario for those systems unlikely.

Large disk sizes (i.e. $r_\mathrm{Hill}/3$) indicate the formation in fragmenting disks or molecular cores. For such a scenario the disk evolution might be similar to disks around low-mass stars. For example, \citet{Stamatellos2015} argued that the disks of PMCs formed through fragmentation of the parent disk should be more massive than expected from scaling relations derived from disks around low-mass stars (i.e. $M_\mathrm{d}>1\%$ of the PMC mass; see also \citealt{Wu2017b}). They also argued that the CPDs evolve independently after they have separated from their parent disk and might be long-living owing to their initially high mass. \citet{Schwarz2016a} for \mbox{GQ Lup b} and \citet{Ginski2018} for \mbox{CS Cha c} argued that their derived high eccentricities are not compatible with the in situ formation in the parent disk but point towards the formation within the molecular cloud, similar to binary formation. In both cases the companions should have disks that evolve similar to disks around low-mass stars. Such a scenario is also supported by spin measurements of wide-orbit PMCs and brown dwarfs, showing that they follow a very similar spin-evolution and spin-mass relation. This indicates that the spin of these objects is regulated by their surrounding accretion disk \citep{Bryan2018,Scholz2018}. We therefore would expect a detection with the observing strategy proposed in this work at least for the more massive candidates. Companions with highly inclined or high eccentric orbits and a detection of a gas disk would be a strong argument for the fragmenting molecular cloud (binary formation) scenario.
\subsection{Impact of the primary's disk}
\label{sec:embeddedpmc}
For all our models, we assumed that the CPD of the PMC is not affected by the protoplanetary disk of the host star and that the observables of the CPD are not affected by any kind of background emission. Such background emission would certainly make the direct detection of a CPD harder or even unlikely. 

If the companion is embedded in the protoplanetary disk, it  forms a gap during its formation \citep[e.g.][]{Takeuchi1996,Isella2016}, which might allow us to directly detect embedded CPDs. \citet{Szulagyi2018} have shown that it is feasible to detect CPDs in gaps with ALMA continuum observations, assuming that the disk is seen face-on. For the gas, the situation is more complicated as the velocity field has to be taken into account. As shown by models of \citet{Perez2018} and indicated by the observations of \citet{Pinte2018}, an embedded planet disturbs the local Keplerian velocity field of the protoplanetary disk. The gas emission of the companions CPD is on top of this disturbed emission, which makes the separation of the CPD signal from the protoplanetary disk signal very challenging. However, a detection of the embedded CPD in the gas might be feasible as shown by \citet{Perez2015a}. 

To directly detect a still-embedded CPD most likely requires higher spatial ($<0.1\arcsec$) and spectral ($<1.0\,\mathrm{km\;s^{-1}}$) resolution as we used, for example, for the CS~Cha companion. That makes a direct detection of an embedded CPD in the gas very unlikely, considering similar observing times as used in this work. This is also indicated by the recent ALMA high spatial resolution ($\approx0.\arcsec035$) survey of 20 protoplanetary disks \citep[DSHARP; ][]{Andrews2018}. For example, the planet-induced kink in the velocity field of the \mbox{HD 163296} disk, reported by \citet{Pinte2018}, is also seen in the high spatial resolution $^{12}\mathrm{CO}$ DSHARP observations, but a more detailed investigation of the kink is not feasible \citep{Isella2018} as a consequence of the lower spectral resolution compared to the \citet{Pinte2018} data. This shows the necessity for high spatial and spectral resolution line observations to detect embedded CPDs, which is challenging even with ALMA. Nevertheless, such surveys are ideal to study observational signatures of planet disk interaction. In any case, a detailed study of embedded CPDs requires complex three-dimensional modelling, which is out of the scope of this paper. In the context of any residual emission from the disk of the primary, our results should be considered as a \mbox{best-case} scenario.

As discussed in Sect.~\ref{sec:formationscenario}, wide-orbit PMCs might not be formed in the protoplanetary disk of their host star at all. In such a case, the orbit of the PMC is likely not coplanar with the protoplanetary disk and it is possible to have an undisturbed view towards the CPD. That depends on the orbit geometry of the system and on the location of the PMC at the time  of the observations (i.e. the PMC can be in front or behind the protoplanetary disk). One such system might be GQ~Lup. The observation of \citet{MacGregor2017} have indicated that the PMC is located inside the gaseous protoplanetary disk, but no disturbances in the velocity field were found. On the other hand \citet{Schwarz2016a} argued that a highly inclined orbit of \mbox{GQ Lup b} is also possible. In this case, higher spatial ALMA observations are required to confirm that the PMC is indeed embedded in the protoplanetary disk.  

Another interesting example is the former PMC candidate in the FW Tau system, although it turned out to be a low-mass stellar object. The gas observations of \citet{Wu2017a} have shown that the protoplanetary disk around the central close binary already dissipated, whereas the companion \mbox{FW Tau c} still hosts a prominent dust and gas disk. This indicates that the circumbinary disk might have dissipated on a shorter timescale than the disk around FW Tau c \citep{Wolff2017}. 

Such a scenario also seems likely for the CS~Cha system. The VLT/SPHERE scattered light images indicate a circumbinary dust disk with a size of $r\approx 56\,\mathrm{au}$, which is significantly smaller than the orbital distance $a\approx214\,\mathrm{au}$ of the companion detected by \citet{Ginski2018}. Compared to (sub)mm observations, the VLT/SPHERE observations trace smaller grains ($\lesssim 1\,\mathrm{\mu m}$) that are well coupled to the gas and therefore provide a better estimate of the gas disk extension. The fact that CS~Cha is a close-binary system makes it a very interesting candidate for a CPD detection experiment.
\subsection{Moon formation}
The regular moons of Jupiter, Saturn, and Uranus are thought to have formed in CPDs as a consequence of planet formation \citep{Lunine1982,Pollack1974a,Canup2002, Mosqueira2003,Szulagyi2018a} or from the spreading of massive rings \citep{Crida2012}. A minimum mass Jovian CPD analogous to the Hayashi Minimum Mass Solar Nebulae \citep{Hayashi1981} would have a mass of $\approx0.023\,M_\mathrm{J}$ and a radius of $\approx0.014\,\mathrm{au,}$ which results in an average surface density of $\Sigma \approx 10^5\,\mathrm{g\,cm^{-2}}$ \citep{Lunine1982}. As our results show, such a compact and dense disk would not be detectable with (sub)mm telescopes. Nevertheless we use the Jovian CPD as a reference to discuss moon formation in context of our CPD models. 

The mass of this Jovian CPD is actually very similar to the mass of the CS Cha c CPD model presented in Sect.~\ref{sec:cscha}, implying that the CS Cha c CPD can still form moons similar to the Galilean moons of Jupiter. However, in the Jovian CPD the type I migration timescale is shorter than the disk Kelvin-Helmholtz cooling timescale, preventing the formation of massive or volatile-rich satellites \citep{Canup2002}. Furthermore, dust evolution models suggest very rapid dust depletion (see Sect.~\ref{sec:results_dust}) of such compact and dense disks, which might also prevent moon formation.

\citet{Canup2002} proposed a accretion disk model with about a 1000 times lower disk mass than the minimum mass Jovian CPD. In such a disk, satellites are lost more slowly to inward migration, but it requires a continuous supply of material into the CPD to allow for massive satellites to be present at the time of disk dispersal \citep{Pollack1974,Canup2009}. In such a scenario the CPDs of isolated wide-orbit PMCs might not be able to form a Galilean moon system owing to the lack of material to replenish the disk. Alternatively the moons might have already formed before those CPDs became isolated. In that case we would expect very low dust-to-gas mass ratios, something that can be tested with observations as suggested in this work.

Alternatively the CPD ice line, dead zone, or the presence of an inner cavity have been suggested as mechanisms to prevent satellite loss via migration \citep{Sasaki2010,Fujii2014,Heller2015}. With our models it is actually possible to determine the location of the ice lines and dead zones as chemical processes are also included. Such models combined with constraints from observations can therefore provide important input for moon formation theories.
\subsection{PMCs versus free-floating planet-mass objects}
In contrast to PMCs millimetre-dust emission was already detected around the free-floating planet-mass object OTS 44 \citep{Bayo2017}. There are many more free-floating planet candidates \citep[see][]{Caballero2018} but so far only a few of these candidates show indications of a CPD \citep[see][]{Bayo2017}. For our presented modelling approach the only significant difference compared to PMCs is that the size of CPDs around free-floating planets is not limited by the Hill radius as they do not orbit a star or brown dwarf. The planetary properties of OTS 44 $L_\mathrm{p}\approx2.4\times10^{-3}\,L_\mathrm{\sun}$, $M_\mathrm{p}\approx 6-14\,M_\mathrm{J}$ \citep{Joergens2013,Bayo2017} and the estimated range for the disk mass $M_\mathrm{d}\approx 0.02-0.2\,M_\mathrm{J}$ (\citealt{Bayo2017}, assuming $d/g=0.01$) are within the parameter space covered in this work. Therefore, our results are also valid for \mbox{OTS 44} and other free-floating planets, in particular the detection limits for the gas component. Spatially resolved observations can reveal if CPDs around wide-orbit PMCs are indeed smaller than disks around free-floating planets and provide constraints for disk formation and evolution (e.g. viscous spreading) theories.
\section{Conclusions}
\label{sec:conclusions}
We have presented radiation thermochemical models for CPDs around PMCs ($M_\mathrm{p}\lesssim20\,M_\mathrm{J}$) on wide orbits ($a\gtrsim 100 \,\mathrm{au}$). We assumed that those companions are already separated from or were not formed at all in the disk of their host star and that observations of the CPD are not affected by the presence of a protoplanetary disk. For these self-consistent dust and gas models, we produced synthetic observables for both the gas and dust to investigate the potential of (sub)mm telescopes to directly detect CPDs. We compared our results to sensitivity estimates for ALMA and the future ngVLA at various frequencies, assuming $\approx6\,\mathrm{h}$ of on-source integration time. Our main conclusions are as follows:\begin{enumerate}
  \item Isolated CPDs are detectable in $^{12}\mathrm{CO}$ spectral lines if they are large enough. For objects at distances of $d\approx150\,\mathrm{pc}$ the outer disk radius needs to be larger than $r_\mathrm{out}\gtrsim10\,\mathrm{au}$, if the lines are optically thick throughout the disk. Assuming that the CPD outer radius is one-third of the companions Hill sphere, this implies that a CPD around a $M_\mathrm{p}=10\,M_\mathrm{J}$ planet orbiting a solar-mass star on an orbit $a\gtrsim200\,\mathrm{au}$ is detectable even if its disk mass is as low as $\approx10^{-4}\,M_\mathrm{J}$ (see Fig.~\ref{fig:hillradii}).
  \item Similar to the analytic approach of \citet{Wu2017}, we find that the dust disks of such CPDs need to be small ($r\lesssim2\,\mathrm{au}$) to match the current upper limits of ALMA continuum observations and to be consistent with the VLT/SPHERE observations of CS Cha c \citep{Ginski2018}. However, dust evolution, such as rapid radial migration, can lead to strongly dust depleted disks ($d/g\lesssim10^{-3}$) on short timescales ($\lesssim10^5\,\mathrm{yr}$; see also \citealt{Zhu2018}). In that case the gas disk can still be as large as one-third of the Hill radius, and such a disk is detectable. 
   \item The best frequencies for such observations are either ALMA Band 7 or Band 6 of the future ngVLA. However, several factors such as the beam size, spectral resolution, the disk/planet properties (i.e. luminosity, inclination ) and their environment (i.e. radiation from the stellar host) need to be considered. Models such as presented in this work are therefore crucial to optimize observing programmes for gas and dust observations of wide-orbit CPDs.
\end{enumerate}
Gas observations of CPDs around wide-orbit companions will provide crucial information on their formation scenario and for planet formation theories in general. Known wide-orbit PMCs at distances of $140-160\,\mathrm{pc}$ that show accretion tracers should be detectable with deep ALMA observations of $^{12}\mathrm{CO}$ spectral lines because the size of their Hill sphere allows for disk sizes larger than our derived detection limit of $r_\mathrm{out} \approx10\,\mathrm{au}$. They are therefore the ideal targets to perform detection experiments for gaseous CPDs. Even in case of non-detections, such deep observations will allow us to determine stringent upper limits on the disk size and gas mass. The detection of a Keplerian rotation signature would provide clear evidence that those planet-mass objects indeed host a rotating disk and would likely provide a more accurate estimate of the companion mass than optical/near-infrared photometric observations. This would answer the question of whether those wide-orbit companions are really planets. 
\begin{acknowledgements}
We want to thank the anonymous referee for a very constructive report that improved the paper. C. R., G. M.-A., and C. G. acknowledge funding from the Netherlands Organisation for Scientific Research (NWO) TOP-1 grant as part of the research programme “Herbig Ae/Be stars, Rosetta stones for understanding the formation of planetary systems”, project number 614.001.552. This research has made use of NASA's Astrophysics Data System. All figures were made with the free Python module Matplotlib \citep{Hunter2007}. This research made use of Astropy, a community-developed core Python package for Astronomy \citep{AstropyCollaboration2013}. We would like to thank the Center for Information Technology of the University of Groningen for their support and for providing access to the Peregrine high performance computing cluster.
This work has made use of data from the European Space Agency (ESA) mission
{\it Gaia} (\url{https://www.cosmos.esa.int/gaia}), processed by the {\it Gaia}
Data Processing and Analysis Consortium (DPAC,
\url{https://www.cosmos.esa.int/web/gaia/dpac/consortium}). Funding for the DPAC has been provided by national institutions, in particular the institutions
participating in the {\it Gaia} Multilateral Agreement.
\end{acknowledgements}
%%%%%%%%%%%%%%%%%%%%%%%%%%%%%%%%%%%%%%%
\bibliographystyle{aa}
\bibliography{isocpd}
%%%%%%%%%%%%%%%%%%%%%%%%%%%% 
\begin{appendix} %First appendix
\section{Companion spectrum}
\label{sec:planetaryspectra}
To simulate the irradiation of the disk by the PMC, we use \mbox{DRIFT-PHOENIX} brown dwarf/giant planet atmosphere models \citep{Witte2009,Witte2011} to construct the photospheric spectrum. To account for possible accretion and the resulting accretion luminosity $L_\mathrm{accr}$, we consider the luminosity produced by a shock at the planetary surface due to infalling material. For this emission component, we use a black-body spectrum with $T=10000\,\mathrm{K}$ and assume that the emission is distributed over the whole surface and that all the released energy is radiated away. The accretion luminosity is then related to the accretion rate via \citep[e.g.][]{Zhu2015,Marleau2017a} 
\begin{equation}
L_\mathrm{accr}=\frac{GM_{p}\dot{M}_\mathrm{accr}}{R_\mathrm{p}},
\end{equation} 
where $G$ is the gravitational constant, $M_{p}$ is the mass of the planet, $\dot{M}_\mathrm{accr}$ is the mass accretion rate, and $R_\mathrm{p}$ is the radius of the planet.
\section{Temperature structure for the reference model}
\label{sec:reftemp}
\begin{figure}
\includegraphics[width=\hsize]{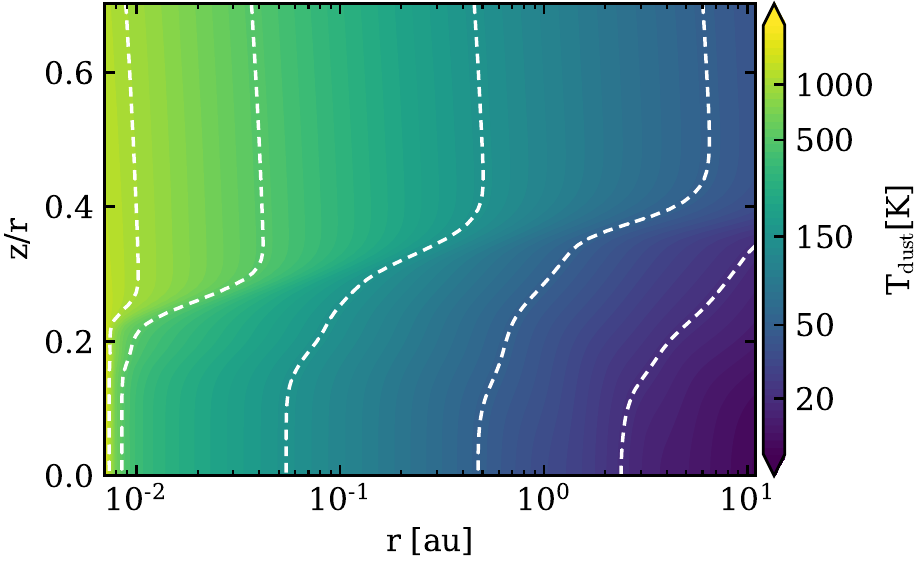}\\
\includegraphics[width=\hsize]{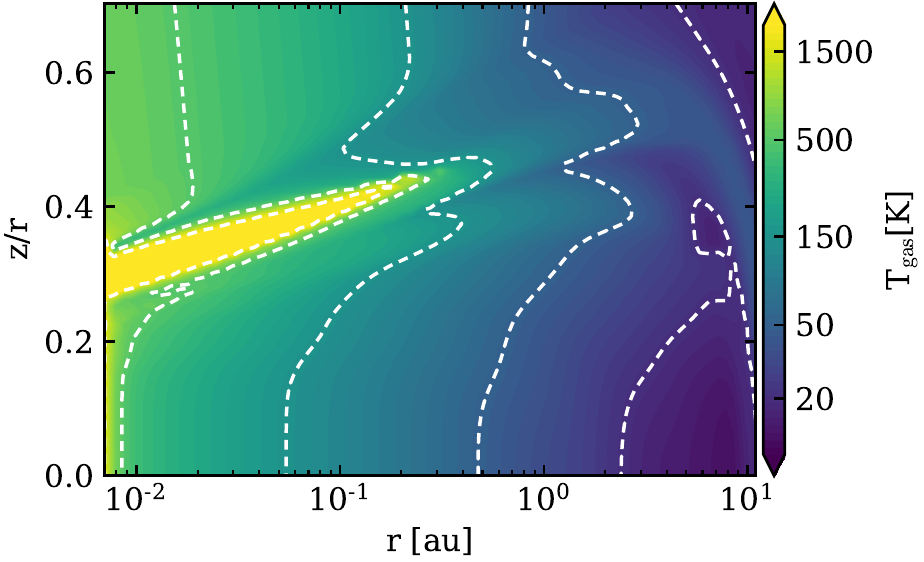}
\caption{Dust (top panel) and gas (bottom panel) temperature structure for the reference model (see Sect.~\ref{sec:methodrefmodel}). The white dashed contour lines correspond to the temperature values given in the colour bar. The logarithmic colour scale is the same in both plots.}
\label{fig:reftemp}
\end{figure}

In Fig.~\ref{fig:reftemp} we show the resulting dust and gas temperature structure for the reference model (see Sect.~\ref{sec:methodrefmodel}). The minimum, maximum, and mass averaged temperatures for the dust are $T_\mathrm{d,min}=10.7\,\mathrm{K}$, $T_\mathrm{d,max}=1550.9\,\mathrm{K,}$ and $T_\mathrm{d,avg}=57.8\,\mathrm{K}$; these temperatures for the gas are $T_\mathrm{g,min}=12.4\,\mathrm{K}$, $T_\mathrm{g,max}=4353\,\mathrm{K,}$ and $T_\mathrm{d,avg}=57.8\,\mathrm{K}$. The region with $T_\mathrm{g}>1500\,\mathrm{K}$ is mainly due to heating by the accretion luminosity of the PMC.
\section{Sensitivity calculations}
\label{sec:almasens}
\begin{table*}
\caption{Sensitivity (rms) values for the considered ngVLA and ALMA bands/lines.}
\label{table:sensitivity}
\centering          
\begin{tabular}{l c c c c c c}
\hline\hline
Line & Wavelength & Frequency & Band & Line rms & Continuum rms & Beam (FWHM) \\
 & $\mathrm{[\mu m]}$ & GHz & & $\mathrm{[Jy/beam]}$ & $\mathrm{[Jy/beam]}$ & mas \\
\hline
$\mathrm{CO\,J=1-0}$ & 2600.76 & 115.27 & ngVLA B6 & 7.9(-5) & 3.1(-7) & 4 \\
$\mathrm{CO\,J=2-1}$ & 1300.40 & 230.54 & ALMA B6  & 4.6(-4) & 4.7(-6) & 17 \\
$\mathrm{CO\,J=3-2}$ & 866.96  & 345.80 & ALMA B7  & 6.5(-4) & 8.1(-6) & 12 \\
$\mathrm{CO\,J=4-3}$ & 650.25  & 461.04 & ALMA B8  & 2.1(-3) & 3.0(-5) & 9 \\
$\mathrm{CO\,J=6-5}$ & 433.56  & 691.47 & ALMA B9  & 6.8(-3) & 1.2(-4) & 6 \\
$\mathrm{CO\,J=7-6}$ & 371.65  & 806.65 & ALMA B10 & 1.5(-2) & 2.9(-4) & 5 \\
\hline
\end{tabular}
\end{table*}
We used the ALMA sensitivity calculator\footnote{\url{https://almascience.eso.org/documents-and-tools/proposing/sensitivity-calculator}; Version from 04.10.2018} to determine the sensitivities for the line and continuum observations in the various bands. We used an on-source integration time of $6\,\mathrm{h}$ ($\approx12\,\mathrm{h}$ with overheads) using $50$ antennas and best weather conditions (PWV value of $0.472\,\mathrm{mm}$). For the spectral lines we used a bandwidth of $1\,\mathrm{km\,s^{-1}}$ and for the continuum a bandwidth of $\mathrm{7.5\,GHz}$.

For the $2600\,\mathrm{\mu m}$ emission ($\mathrm{CO}\,J\!=\!1\!-\!0$), we did not use ALMA as reference but the ngVLA because the expected sensitivity for the ngVLA is about an order of magnitude better than for ALMA Band~3. The sensitivity values are taken from \url{https://science.nrao.edu/futures/ngvla/concepts} (version from 04.10.2018). We chose the values for a beam size of $4\,\mathrm{mas}$ (from the Table with Natural Weighting) and scaled these to the observing time of $6\,\mathrm{h}$ and for the lines to a bandwidth of $1\,\mathrm{km\,s^{-1}}$. The resulting rms values together with the maximum spatial resolution for all bands/lines considered are listed in Table~\ref{table:sensitivity}.
\section{ALMA/CASA simulations}
\label{sec:almasim}
We produced spectral line cubes for several $^{12}\mathrm{CO}$ transitions  using the line transfer module of \prodimo \citep{Woitke2011}. The model line cubes have a spectral resolution of $0.25\,\mathrm{km/s}$. To determine if those models are actually observable with ALMA or ngVLA, we convolved the model line cubes with a Gaussian beam of the desired size and binned the channels to the spectral resolution (channel width) used to determine the sensitivity limits. For this, we used the \texttt{image} tools and the task \texttt{specsmooth} of the CASA software (Common Astronomy Software Applications, Version 5.3.0-143; \citealt{McMullin2007}). Contrary to full ALMA/CASA simulations, this approach is very efficient and fast as it does not require the simulation of visibilities. This allows us to explore several lines and many models. This approach is used for the results shown in Fig.~\ref{fig:cstruc} and all other similar figures.

For selected models we also produced realistic ALMA simulations using the CASA tasks \texttt{simobserve} and \texttt{simanalyze}. For \texttt{simobserve} we used a \texttt{totaltime} of $6\,\mathrm{h}$ and simulated the noise (\texttt{thermalnoise=tsys-atm}) assuming a perceptible water vapour column of \texttt{user\_pwv=0.5} (i.e. very good weather conditions). We binned the simulated line cube to a spectral resolution of $1\,\mathrm{km\,s^{-1}}$ prior to the cleaning. With this configuration we reached rms noise levels in the individual channels similar to the values derived via the sensitivity calculator (see Appendix~\ref{sec:almasens}). The synthetic images were produced with the \texttt{simanalyze} task using natural weighting and a threshold close to the average rms value. The results (i.e. peak fluxes) of the full simulations are consistent within a few percent with the results from the simpler, but faster, beam convolution approach. 
\section{Radial column density profiles}
\begin{figure}
\includegraphics[width=\hsize]{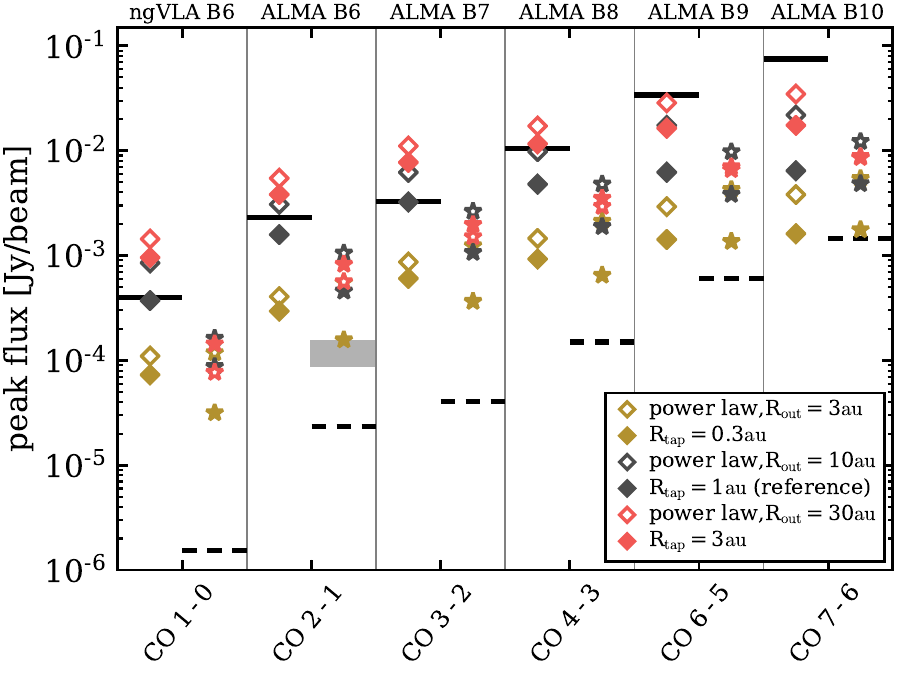}
\caption{Same as Fig.~\ref{fig:cstruc} but for models with \mbox{tapered-edge} and power-law radial column density profiles.}
\label{fig:cstrucCD}
\end{figure}
\begin{figure}
\centering
\resizebox{\hsize}{!}{\includegraphics{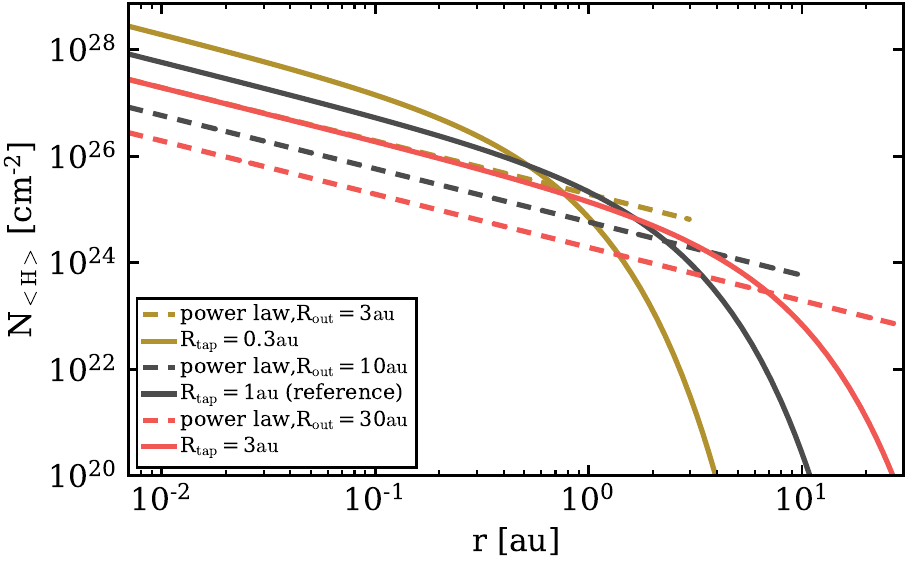}}\\
\caption{Radial column density profiles for the models shown in Fig.~\ref{fig:cstrucCD}.}
\label{fig:cStrucCDNH}
\end{figure}
\begin{figure}
\centering
\resizebox{\hsize}{!}{\includegraphics{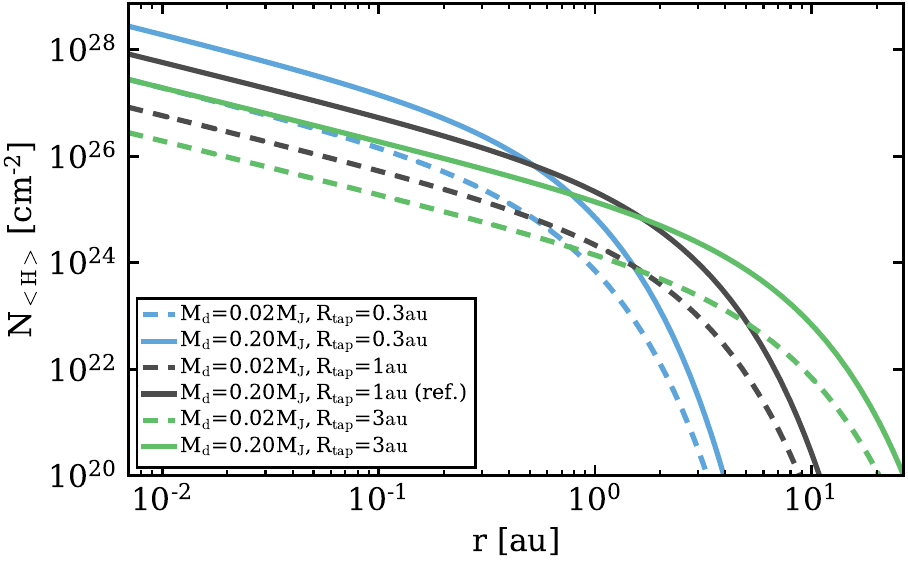}}\\
\caption{Radial column density profiles for the models shown in Fig.~\ref{fig:cstruc}.}
\label{fig:cStrucNH}
\end{figure}

For our models, we mainly use disk structures with a \mbox{tapered-edge} radial column density profile. In Fig.~\ref{fig:cstrucCD} we compare pure power-law radial column density models to \mbox{tapered-edge} models, where all models have the same disk mass. The figure shows that the general picture is not affected by the type of profile used. The power-law models are brighter than their corresponding \mbox{tapered-edge} counterpart. This is mainly driven by their slightly larger outer disk radius (see Fig.~\ref{fig:cStrucCDNH}) and their higher column density at the outer edge of the disk. Nevertheless, the disk size remains the main driving factor for both the dust and gas emission.
In Fig.~\ref{fig:cStrucNH} we also show the radial column density profiles for the models with varying disk extension and disk masses. These are the same models as shown in Fig.~\ref{fig:cstruc}.
\section{Dust evolution in circumplanetary disks}
\label{sec:dustevol}
\begin{figure*}
\centering
\includegraphics{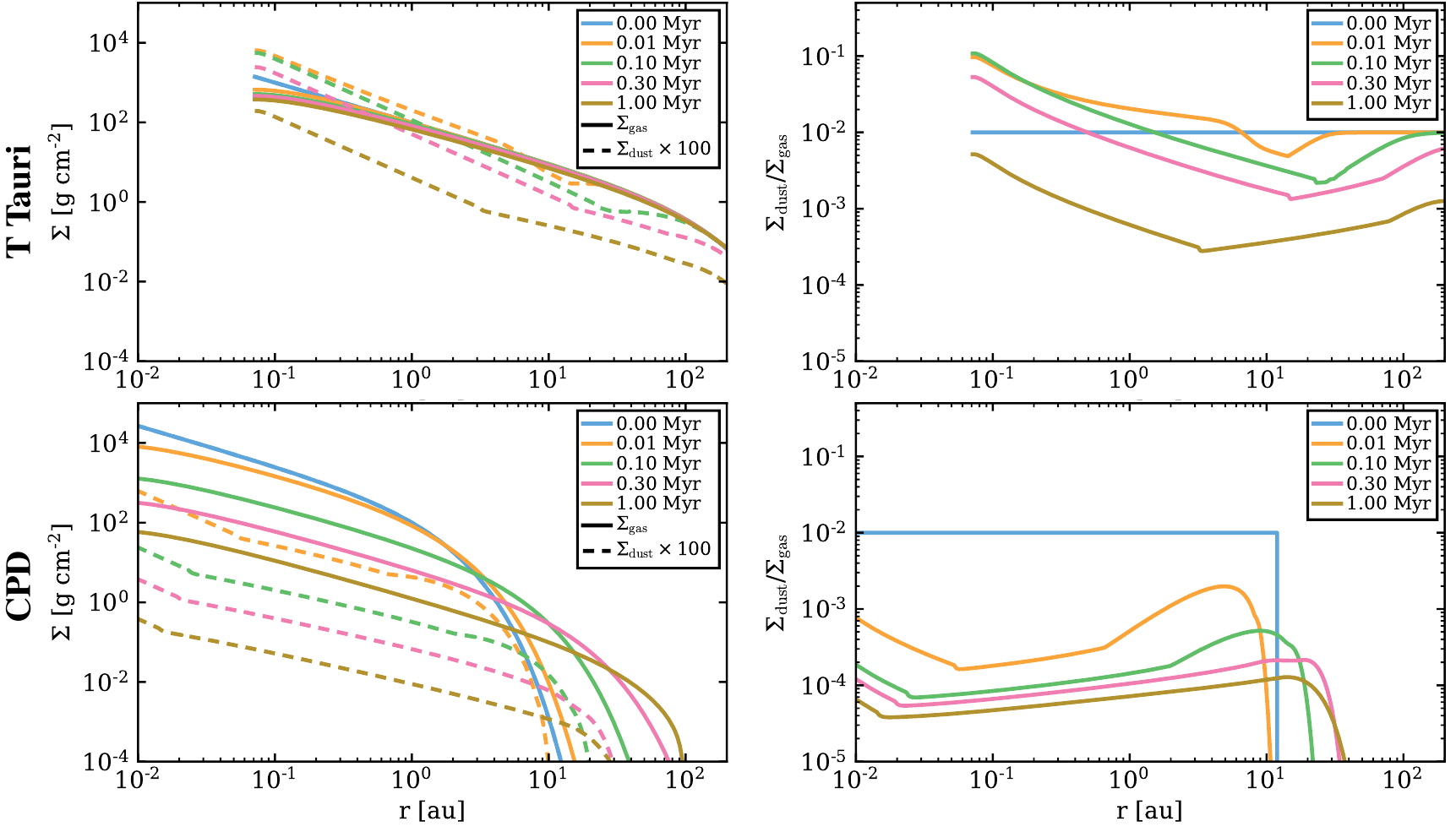}
\caption{Results from the \emph{two-pop-py} dust evolution model. The top row represents a T~Tauri disk model; the bottom row shows the reference CPD model. The left column shows the time evolution of the gas (solid lines) and dust (dashed lines) surface densities as a function of radius; the right column shows the evolution of the dust-to-gas mass ratio.}
\label{fig:twopoppy}
\end{figure*}
Similar to protoplanetary disks, the dust in CPDs might evolve through growth, fragmentation, and radial migration. However, timescales for such processes in small compact disks might be very different. \citet{Zhu2018} estimated that the drift timescale in CPDs for millimetre-sized grains can be up to three orders of magnitudes smaller compared to T~Tauri disks and only of the order of $100-1000\,\mathrm{yr}$. Also the dust-evolution models of \citet{Pinilla2013} indicated that dust evolution in compact disks around brown dwarfs is faster than for T~Tauri stars and that brown-dwarf disks would be dust poor already after $\lesssim1.5~\mathrm{Myr}$, if there are no efficient mechanisms to stop radial inward migration.

To get a glimpse of the dust evolution in CPDs, we used the \emph{two-pop-py}\footnote{\url{https://github.com/birnstiel/two-pop-py} \mbox{Version: bd3c2552cfd4008e6201b9d8189cf598dfcacfad}} dust evolution code of \citet{Birnstiel2012}. We used our CPD model parameters (see Table \ref{table:discmodel}) as input parameters for \emph{two-pop-py}, but also did run a reference T~Tauri disks model with the physical parameters from \citet{Woitke2016}. All other parameters for the \emph{two-pop-py} code such as the turbulence alpha value ($\alpha=10^{-3}$) or the efficiency for sticking and drift (both are unity) are the same in both models. 

Fig.~\ref{fig:twopoppy} shows a comparison of the time evolution of the dust and gas surface density and the dust-to-gas mass ratio of those two models. In the CPD model, the dust-to-gas mass ratio drops to values of $d/g\lesssim10^{-3}$ already after $10^4\,\mathrm{yr}$ and to values of $d/g\lesssim10^{-4}$ after $1~\mathrm{Myr}$, whereas for the T~Tauri disk it takes $1~\mathrm{Myr}$ to reach $d/g\lesssim10^{-3}$. This indicates that the radial drift in CPDs happens about 100 times faster than in T~Tauri disks, which is consistent with the estimates of \citet{Zhu2018}. We note that for the lower mass-accretion rates the radial migration would be even faster (see also \citealt{Zhu2018}).
\section{Synthetic channel maps for CS Cha c}
\label{sec:cschaalma_more}
\begin{figure*}
\includegraphics[width=\hsize]{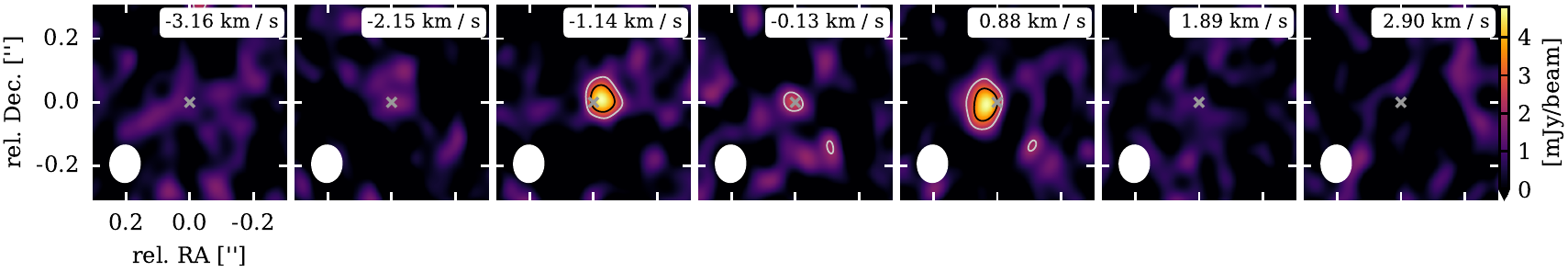}
\includegraphics[width=\hsize]{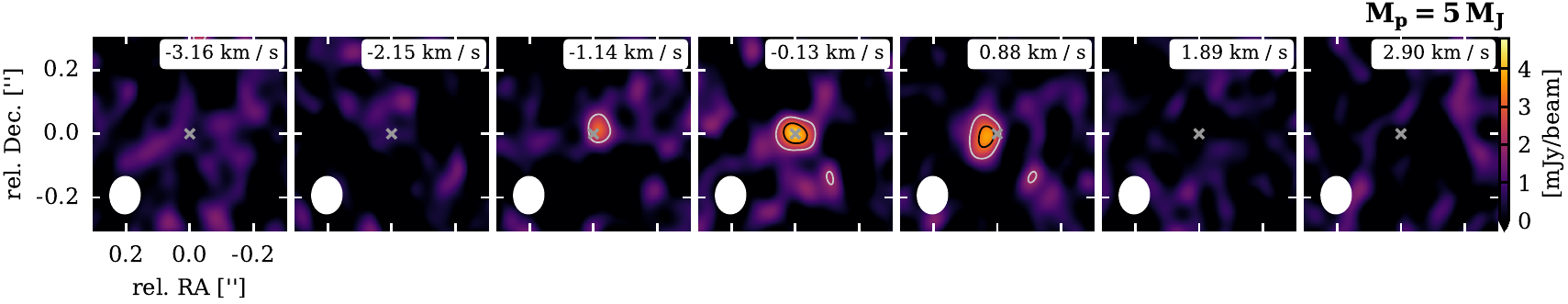}
\includegraphics[width=\hsize]{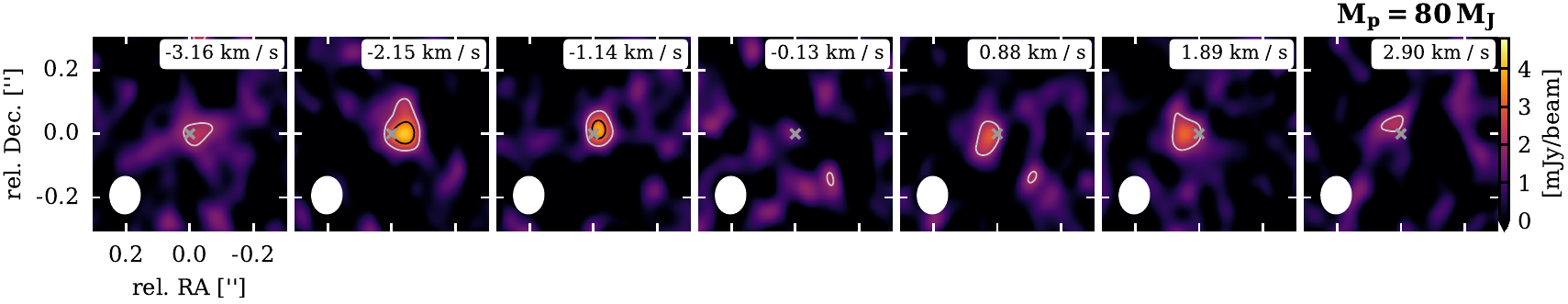}
\includegraphics[width=\hsize]{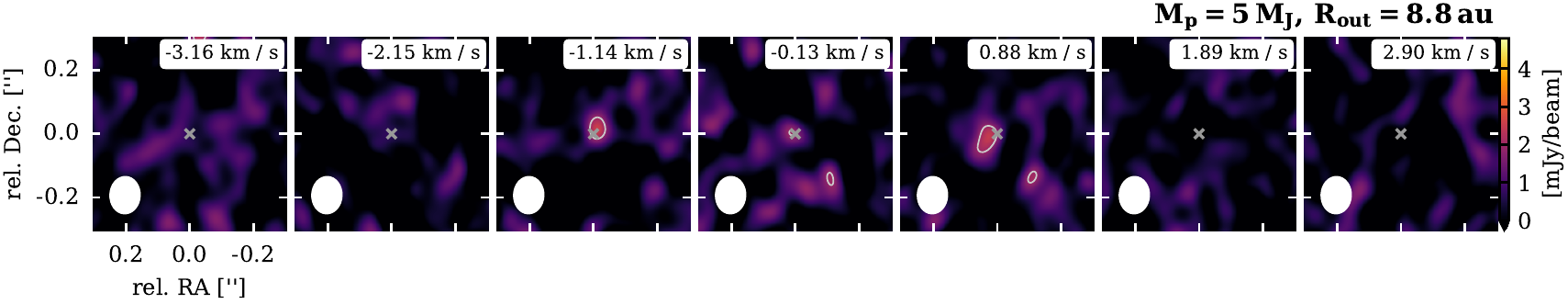}
\includegraphics[width=\hsize]{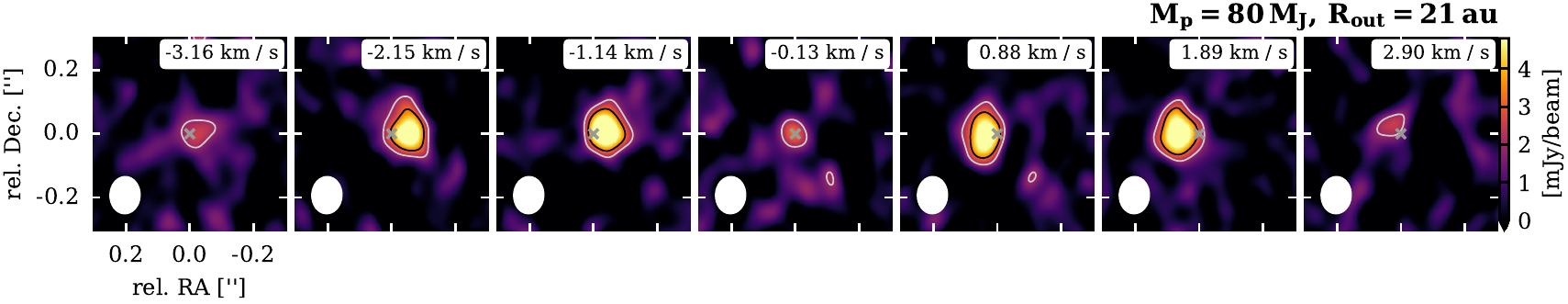}
\caption{$^{12}\mathrm{CO}\,J\!=\!3\!-\!2$ (ALMA Band 7) synthetic channel intensity maps for CS~Cha companion models. The grey cross indicates the central target location. The white ellipse in each panel shows the beam with a size of $0.11\arcsec\times0.09\arcsec$. In the top right corner of each panel the  velocity relative to the systemic velocity is indicated. The grey and black contours show 3 and 5 times the theoretical rms level of $0.65~\mathrm{mJy/beam}$ (see Table~\ref{table:sensitivity}). In the top row we show the same model as in Fig.~\ref{fig:cschaalma} (Sect.~\ref{sec:cscha}) for easier comparison. The second and third row show models where only the PMC mass was changed. For the models in the last two rows the disk outer radius was adapted to $r_\mathrm{Hill}/3$ according to the PMC mass. The spatial and colour scale are the same for all five plots.}
\label{fig:cschaalma_more}
\end{figure*}

In Fig.~\ref{fig:cschaalma_more} we show $^{12}\mathrm{CO}\,J\!=\!3\!-\!2$ synthetic channel maps for the CS~Cha companion model (see Sect.~\ref{sec:cscha}) with varying PMC mass $M_\mathrm{p}$ and outer disk radius $R_\mathrm{out}$. All other parameters are identical to the model shown in Fig.~\ref{fig:cschaalma}. The mass of the planet influences the disk velocity field and therefore the spectral line emission. In Fig.~\ref{fig:cschaalma_more} we show the impact of the PMC mass on the channel maps assuming the same ``observing'' set-up as was used in the model presented in Fig.~\ref{fig:cschaalma}. We also applied the exact same procedure (see Appendix~\ref{sec:almasim}) for producing the synthetic channel maps to all the models; for example, we did not optimize the cleaning process.

Two different types of models are shown in Fig.~\ref{fig:cschaalma_more}. For the $M_\mathrm{p}=5\,M_\mathrm{J}$ and $M_\mathrm{p}=80\,M_\mathrm{J}$ only the companion mass was changed; the disk density, temperature, and chemical structure were fixed; i.e. we only allowed for a change in the velocity field. For the other two models, the disk radius was set according to the PMC mass to $r_\mathrm{Hill}/3$. All other properties of the PMC (i.e. luminosity) and the CPD (i.e. disk mass) are the same in all the shown models. For the $M_\mathrm{p}=5\,M_\mathrm{J}, R_\mathrm{out}=8.8\,\mathrm{au}$ model the disk is too small to be clearly detected.
\end{appendix}
\end{document}